\documentclass[11pt]{article}
\usepackage{amsmath}

\usepackage{times}

\usepackage{moreverb}  % for boxedverbatim (which shouldn't be used in lists)

\begin{document} \sloppy
\pagenumbering{roman}
\setcounter{page}{1}
\thispagestyle{empty}
\begin{center}
% \vspace*{-1in}
Argonne National Laboratory \\
9700 South Cass Avenue\\
Argonne, IL 60439

\vspace{.2in}
\rule{1.5in}{.01in}\\ [1ex]
ANL/MCS-TM-263 \\
\rule{1.5in}{.01in}

\vspace{1in}
{\Large\bf O{\large\bf TTER} 3.3 Reference Manual}

\vspace{.2in}
by \\ [3ex]

{\Large\it William McCune}\\
%%%%%%%%%%%%%%%%%%%%%
\thispagestyle{empty}

\vspace{1.5in}
Mathematics and Computer Science Division

\bigskip

Technical Memorandum No. 263

\vspace{1in}
August 2003
\end{center}

\vfill
\noindent
\emph{
This work was supported by the Mathematical, Information, and Computational 
Sciences Division subprogram of the Office of Advanced Scientific Computing 
Research, Office of Science, U.S. Department of Energy, under Contract 
W-31-109-ENG-38.}
%%%%%%%%%%%%%%%%%%%%%

%%% Local Variables: 
%%% mode: latex
%%% TeX-master: "snadiopt"
%%% End: 
\newpage
\pagenumbering{roman}
\setcounter{page}{2}
\noindent
Argonne National Laboratory, with facilities in the states of Illinois
and Idaho, is owned by the United States Government and operated by The
University of Chicago under the provisions of a contract with the
Department of Energy.

\vspace{2in}

\begin{center}
{\bf DISCLAIMER}
\end{center}

\noindent
This report was prepared as an account of work sponsored by an agency
of the United States Government.  Neither the United States Government
nor any agency thereof, nor The University of Chicago, nor any of
their employees or officers, makes any warranty, express or implied,
or assumes any legal liability or responsibility for the accuracy,
completeness, or usefulness of any information, apparatus, product, or
process disclosed, or represents that its use would not infringe
privately-owned rights.  Reference herein to any specific commercial
product, process, or service by trade name, trademark, manufacturer,
or otherwise, does not necessarily constitute or imply its
endorsement, recommendation, or favoring by the United States
Government or any agency thereof.  The views and opinions of document
authors expressed herein do not necessarily state or reflect those of
the United States Government or any agency thereof.
\newpage
  \pagestyle{plain}
  \tableofcontents
\newpage

%%%%%%%%%%%%%%%%%%%%%%%%%%%%%%%%%%%%%%%%%%%%%%%%%%%%%%%%%

\newcommand{\otter}{{\sc Otter}}
\newcommand{\maxint}{$\infty$}
\newcommand{\unix}{{\sc unix}}

\pagenumbering{arabic}
\setcounter{page}{1}
\title{\otter\ 3.3 Reference Manual}

\author{\emph{William McCune}}

\date{}

\maketitle

\addcontentsline{toc}{section}{Abstract}
\begin{abstract}
\otter\ is a resolution-style theorem-proving program for
first-order logic with equality.  \otter\ includes the inference
rules binary resolution, hyperresolution, UR-resolution, and binary
paramodulation.  Some of its other abilities and features are conversion from
first-order formulas to clauses, forward and back subsumption,
factoring, weighting, answer literals, term ordering, forward and back
demodulation, evaluable functions and predicates, Knuth-Bendix
completion, and the hints strategy.
\otter\ is coded in ANSI C, is free, and is portable
to many different kinds of computer.
\end{abstract}

\section{Introduction}

\otter\ (Organized Techniques for Theorem-proving and Effective
Research) is a resolution-style theorem prover, similar in scope and
purpose to the {\sc aura} \cite{AURA} and {\sc lma/itp} \cite{ITP}
theorem provers, which are also associated with Argonne.
\otter\ applies to statements written in first-order logic with equality.
The primary design considerations have been performance, portability, and
extensibility.  The programming language ANSI C is used.

\otter\ features the inference rules binary resolution,
hyperresolution, UR-resolution, and binary paramodulation.  These
inference rules take a small set of clauses and infer a clause; if the
inferred clause is new, interesting, and useful, it is stored and may
become available for subsequent inferences.
Other features of \otter\ are the following.
\begin{itemize}
\item
Statements of the problem may be input either with first-order
formulas or with clauses (a clause is a disjunction with implicit
universal quantifiers and no existential quantifiers).  If first-order
formulas are input, \otter\ translates them to clauses.
\item
Forward demodulation rewrites and simplifies newly inferred clauses
with a set of equalities, and back demodulation uses a newly inferred
equality (which has been added to the set of demodulators) to rewrite
all existing clauses.
\item
Forward subsumption deletes an inferred clause if it is subsumed by
any existing clause, and back subsumption deletes all clauses that are
subsumed by an inferred clause.
\item
A variant of the Knuth-Bendix method can search for a complete set of
reductions and help with proof searches.
\item
Weight functions and lexical ordering decide the value of
clauses and terms.
\item
Answer literals can give information about the proofs that are found.
See Sec.~\ref{answer}.
\item
Evaluable functions and predicates build in integer arithmetic,
Boolean operations, and lexical comparisons and enable users to
``program'' aspects of deduction processes.
See Sec.~\ref{eval}.
\item
Proofs can be presented in a very detailed form, called
\emph{proof objects}, which can be used by other programs,
for example, to check or to translate the proofs.
See Sec.~\ref{ivy}.
\item
The \emph{hints strategy} can be used to provide heuristic
guidance to the search.  To apply this feature, the user
gives a set of \emph{hint clauses}, and clauses similar to
the hint clauses are emphasized during the search.
See Sec.~\ref{hints}.
\item
\otter's input is compatible, for the most part, with
a complementary program MACE 2.0 \cite{mace2}, which
looks for finite models of first-order statements.  Given
a conjecture, \otter\ can search for a proof, and MACE 
can look for a counterexample, usually from the same
input file.  MACE 2.0 is included in the standard \otter\ 3.3
distribution packages.
\end{itemize}

Although \otter\ has an autonomous mode, most work with \otter\
involves interaction with the user.  After encoding a
problem into first-order logic or into clauses, the user usually
chooses inference rules, sets options to control the processing of
inferred clauses, and decides which input formulas or clauses are to be
in the initial set of support and which (if any) equalities are to be
demodulators.  If
\otter\ fails to find a proof, the user may wish to try again with
different initial conditions.  In the autonomous mode, the user inputs
a set of clauses and/or formulas, and \otter\ does a simple syntactic
analysis and decides inference rules and strategies.  The autonomous mode is
frequently useful for the first attempt at a proof.

\subsection{What \otter\ Isn't}

Some of the first applications that come to mind when one hears
``automated theorem proving'' are number theory, calculus, and
plane geometry, because these are some of the first areas in
which math students try to prove theorems.  Unfortunately,
\otter\ cannot do much in these areas: interesting number theory
problems usually require induction, interesting calculus and analysis
problems usually require higher-order functions, and the first-order
axiomatizations of geometry are not practical.  (Nonetheless, Art
Quaife has proved many interesting theorems in number theory and
geometry using \otter\ 
\cite{quaife-thesis,quaife-geo}.)  For practical theorem proving
in inductive theories, see the work of Boyer, Moore, and Kaufmann
\cite{boyer-moore-2,acl2-approach}.

\otter\ is also not targeted toward synthesizing or verifying formal
hardware or software systems.  See \cite{bm-verify,hw-verify}
for work in those areas.

Summaries of other theorem-proving systems can be found in proceedings
of the recent Conferences on Automated Deduction (CADE) and in
coverage of the CADE ATP System Competition (CASC).

\subsection{History, New Features, and Changes}

There have been several previous releases of \otter, starting with
version 0.9, which was distributed at the 9th International Conference
on Automated Deduction (CADE-9) in May 1988.  Many new features have
been added since then, many bugs have been fixed, and (of course) many
bugs have been introduced.

\subsection{Useful Background}

This manual does not contain an introduction to first-order logic
or to automated deduction.
We assume that the reader knows the basic terminology
including {\it term}
({\it variable}, {\it constant}, {\it complex term}), {\it atom},
{\it literal}, {\it clause}, {\it propositional variable}, {\it function
symbol}, {\it predicate symbol}, {\it Skolem constant}, {\it Skolem
function}, {\it formula}, {\it conjunctive normal form}
({\it CNF}), {\it resolution}, {\it hyperresolution}, and
{\it paramodulation}.
See \cite{chang-lee,loveland,JAR-overview}
for an introductions and overviews of automated theorem proving,
see \cite{siekmann-wrightson,after-25-years}
for collections of important papers, see
\cite{book2} for a list of general problems in the field, and
see \cite{fascinating,kalman-otter,wm-rp:monograph}
for introductions and applications that focus on the use
of \otter.

\section{Outline of \otter's Inference Process} \label{outline}

Once \otter\ gets going with its real work---making
inferences and searching for proofs---it operates on clauses and on
clauses only.  If the user inputs nonclausal first-order formulas,
\otter\ immediately translates them to clauses, by a straightforward
procedure involving negation normal form conversion, Skolemization,
quantifier operations, and conjunctive normal form conversion.

As with its predecessors {\sc aura} and {\sc lma/itp}, \otter's basic
inference mechanism is the {\em given-clause algorithm}, which can be
viewed as a simple implementation of the set of support strategy
\cite{book1a}.  \otter\ maintains four lists of clauses:
\begin{description}
\item[{\tt usable.}]
This list contains clauses that are available to make inferences.
\item[{\tt sos.}]
Clauses in list \verb:sos: (set of support) are not available to make
inferences; they are waiting to participate in the search.
\item[{\tt passive.}]
These clauses do not directly participate in the search; they are used
only for forward subsumption and unit conflict.  The passive list
is fixed at input and does not change during the search.
See Sec.~\ref{passive}.
\item[{\tt demodulators.}]
These are equalities that are used as rules to rewrite newly inferred
clauses.
\end{description}

\noindent
The {\em main loop} for inferring and processing clauses and
searching for a refutation operates mainly on the lists \verb:usable:
and \verb:sos::

{\small
\begin{verbatim}
    While (sos is not empty and no refutation has been found)
        1. Let given_clause be the ``best'' clause in sos;
        2. Move given_clause from sos to usable;
        3. Infer and process new clauses using the inference
              rules in effect; each new clause must have the
              given_clause as one of its parents and members
              of usable as its other parents;  new clauses
              that pass the retention tests are appended to sos;
    End of while loop.
\end{verbatim}
} 

The set of support strategy requires the user to partition the
input clauses into two sets: those with support and those without.
For each inference, at least one of the parents must have support.
Retained inferences receive support.  In other words, no inferences
are made in which all parents are nonsupported input clauses.  At
input time, \otter's list \verb:sos: is the set of supported clauses,
and \verb:usable: is the nonsupported clauses.  (Once the main loop
has started, \verb:usable: no longer corresponds to nonsupported
clauses, because \verb:sos: clauses have moved there.)  \otter's
main loop implements the set of support strategy, because no
inferences are made in which all of the parents are from the initial
\verb:usable: list.

{\em The following paragraph tries to answer the frequently asked
question ``At a certain point, \otter\ has all of the clauses available
to make the inference I want, and one of the potential parents is
selected as the given clause---why doesn't the program make
the inference?''}

\otter's main loop eliminates an important kind of redundancy.
Suppose one can infer clause $C$ from clauses $A$ and $B$, and suppose
both $A$ and $B$ are in list \verb:sos:.  If $A$ is selected as the
given clause, it will be moved to \verb:usable: and inferences
will be made; but $A$ {\em will not} mate with $B$ to infer $C$,
because $B$ is still in \verb:sos:.  We must wait until $B$ has also
been selected as given clause.  Otherwise, we would infer $C$
twice.  (The redundancy would be much worse with inference rules such as
hyperresolution and UR-resolution with which a clause can have
many parents.)  In general,
all parents that participate in an inference must either have been in
the initial \verb:usable: list or have been selected as given clauses.
(This is not true when demodulators are considered as parents.)

\noindent
The procedure for processing a newly inferred clause \verb:new_cl: follows;
steps marked with \verb:*: are optional.

{\small
\begin{verbatim}
   1.  Renumber variables.
 * 2.  Output new_cl.
   3.  Demodulate new_cl (including $ evaluation).
 * 4.  Orient equalities.
 * 5.  Apply unit deletion.
   6.  Merge identical literals (leftmost copy is kept).
 * 7.  Apply factor-simplification.
 * 8.  Discard new_cl and exit if too many literals or variables.
   9.  Discard new_cl and exit if new_cl is a tautology.
 * 10. Discard new_cl and exit if new_cl is too `heavy'.
 * 11. Sort literals.
 * 12. Discard new_cl and exit if new_cl is subsumed by any clause
           in usable, sos, or passive (forward subsumption).
   13. Integrate new_cl and append it to sos.
 * 14. Output kept clause.
   15. If new_cl has 0 literals, a refutation has been found.
   16. If new_cl has 1 literal, then search usable, sos, and
            passive for unit conflict (refutation) with new_cl.
 * 17. Print the proof if a refutation has been found.
 * 18. Try to make new_cl into a demodulator.
   -------------
 * 19. Back demodulate if Step 18 made new_cl into a demodulator.
 * 20. Discard each clause in usable or sos that is subsumed by
            new_cl (back subsumption).
 * 21. Factor new_cl and process factors.
\end{verbatim}
}
%$
\noindent
Steps 19--21 are delayed until steps 1--18 have been applied to all
clauses inferred from the active given clause.

\section{Starting \otter}

Although \otter\ has a primitive interactive feature
(Sec.~\ref{interact}), it is essentially a noninteractive program.  On
{\sc unix}-like systems it reads from the standard input and writes to
the standard output:
\begin{verse}
\tt otter < \it input-file \tt > \it output-file
\end{verse}
No command-line options are accepted; all options are given in the
input file.

\section{Syntax}

\otter\ recognizes two basic types of statement: clauses and formulas.
Clauses are simple disjunctions whose variables are implicitly
universally quantified.  \otter's searches for proofs operate on
clauses.  Formulas are first-order statements without free variables---all
variables are explicitly quantified.  When formulas are input, \otter\
immediately translates them to clauses.

\subsection{Comments}

Comments can be placed in the input file by using the symbol \verb:%:.
All characters from the first \verb:%: on a line to the end of the
line are ignored.  Comments can occur within terms.  Comments are not
echoed to the output file.

\subsection{Names for Variables, Constants, Functions, and Predicates} \label{names}

Three kinds of character string, collectively referred to as
{\em names}, can be used for variables, constants, function
symbols, and predicate symbols:
\begin{itemize}
\item
An {\em ordinary name} is a string of alphanumerics, \verb:$:, and \verb:_:.
%$
\item
A {\em special name} is a string of characters in the set
\verb|*+-/\^<>=`~:?@&!;#| (and sometimes \verb:|:).
\item
A {\em quoted name} is any string enclosed in two
quotation marks of the same type, either \verb:": or \verb:':.  We have
no trick for including a quotation mark of the same type in a quoted name.
\end{itemize}
(The reason for separating ordinary and special names has to do with
infix, prefix, and postfix operators; see Sec.~\ref{ops}.)
For completeness, we list here the meanings of the
remaining printable characters.

\begin{itemize}
\item
\verb:.: (period) --- terminates input expressions.
\item
\verb:%: --- starts a comment (which ends with the end of the line).
\item
\verb:,()[]{}: (and sometimes \verb:|:) --- are punctuation and grouping
symbols.
\end{itemize}

\paragraph{Variables.}

Determining whether a simple term is a constant or a variable depends
on the context of the term.  If it occurs in a clause, the symbol
determines the type: the default rule is that a simple term is a
variable if it starts with \verb:u:, \verb:v:, \verb:w:, \verb:x:,
\verb:y:, or \verb:z:.  If the flag \verb:prolog_style_variables: is
set, a simple term is a variable if and only if it starts with an
upper-case letter or with \verb:_:.  (Therefore, variables in clauses
must be ordinary names.)  In a formula, a simple term is a variable if
and only if it is bound by a quantifier.

\paragraph{Reserved and Built-in Names.}

Names that start with \verb:$:
%$
are reserved for special purposes,
including evaluable functions and predicates (Sec.~\ref{eval}),
answer literals and terms (Sec.~\ref{answer}), and some internal
system names.  The name \verb:=: and any name that starts with
\verb:eq:, \verb:EQ:, or \verb:Eq:, when used as a binary predicate
symbol, is recognized as an equality predicate by the demodulation
and paramodulation processes.  And some names, when they occur
in clauses or formulas, are recognized as logic symbols.

\paragraph{Overloaded Symbols.}

The user can use a name for more than one purpose, for example
as a constant and as a 5-ary predicate symbol.
When the flag \verb:check_arity: is
set (the default), the user is warned about such uses.  Some built-in
names are also overloaded; for example, \verb:|: is used both for
disjunction and as Prolog-style list punctuation, and although the
symbol \verb:-: is built in as logical negation, it can be
used for both unary and binary minus as well.

\subsection{Terms and Atoms} \label{syntax-terms}

Recall that, when interpreted, terms are evaluated as objects in some
domain, and atoms are evaluated as truth values.  Constants and
variables are terms.  An $n$-ary function symbol applied to $n$ terms
is also a term.  An $n$-ary predicate symbol applied to $n$ terms is an
atom. A nullary predicate symbol (also referred to as a
propositional variable) is also an atom.

The pure way of writing complex terms and
atoms is with {\em standard application}: the function or predicate
symbol, opening parenthesis, arguments separated by commas, then
closing parenthesis, for example, \verb:f(a,b,c): and
\verb:=(f(x,e),x):.  If all subterms of a term are written with
standard application, the term is in {\em pure prefix form}.
Whitespace (spaces, tabs, newlines, and comments) can appear in
standard application terms anywhere {\em except} between a function or
predicate symbol and its opening parenthesis.  If the flag
\verb:display_terms: is set, \otter\ will output terms in pure prefix
form.

\paragraph{Infix Equality.}  Some binary symbols can be written in
infix form; the most important is \verb:=:.  In addition, a negated
equality, \verb:-(a=b): can be abbreviated \verb:a!=b:.

\paragraph{List Notation.} Prolog-style list notation can be used
to write terms that usually represent lists.  Table \ref{list-tab} gives
some example terms in list notation and the corresponding pure prefix form.
\begin{table}[htbp] \centering \small
\caption{List Notation}  \label{list-tab}
\begin{tabular}{ll}
\hline
\verb:[]:          &  \verb:$nil: \\
\verb:[x|y]:       &  \verb:$cons(x,y): \\
\verb:[x,y]:       &  \verb:$cons(x,$cons(y,$nil)): \\
\verb:[a,b,c,d]:   &  \verb:$cons(a,$cons(b,$cons(c,$cons(d,$nil)))): \\
\verb:[a,b,c|x]:   &  \verb:$cons(a,$cons(b,$cons(c,x))): \\
\hline
\end{tabular}
\end{table}
%$
Of course, lists can contain complex terms, including other lists.

\subsection{Literals and Clauses}

A literal is either an atom or the negation of an atom.
A clause is a disjunction of literals.
The built-in symbols for negation and disjunction are \verb:-: and \verb:|:,
respectively.
Although clauses can be written in pure prefix form, with \verb:-: as a
unary symbol and \verb:|: as a binary symbol, they are rarely written
that way.  Instead, they are almost always written in infix form, without
parentheses.  For example, the following is a clause in both forms.
\begin{tabbing}
mm\=mmmmmmmmmmmmm\=\kill
\>Pure prefix:         \>  \verb:|(-(a),|(=(b1,b2),-(=(c1,c2)))):\\
\>Infix (abbreviated): \> \verb:-a | b1=b2 | c1!=c2:
\end{tabbing}
\otter\ accepts both forms.  (Clauses are parsed by the general term-parsing
mechanism presented in Sec.~\ref{ops}).

\subsection{Formulas}

Table \ref{ops-tab} lists the built-in logic symbols for constructing
formulas.
\begin{table}[ht] \centering
\caption{Logic Symbols}   \label{ops-tab}
\begin{tabular}{ll}
\hline
negation    & \verb:-: \\
disjunction  & \verb:|: \\
conjunction & \verb:&: \\
implication & \verb:->: \\
equivalence & \verb:<->: \\
existential quantification & \verb:exists: \\
universal quantification   & \verb:all: \\
\hline
\end{tabular}
\end{table}

\paragraph{Formulas in Pure Prefix Form.}

Although the practice is rarely done, formulas can be written in pure
prefix form.  Quantification is the only tricky part: there is a
special variable-arity symbol, \verb:$Quantified:, for quantified
formulas.  For example, $\forall x y \exists z (P(x,y,z) | Q(x,z))$ is
represented by

{\small
\begin{verbatim}
    $Quantified(all,x,y,exists,z,|(P(x,y,z),Q(x,z))).
\end{verbatim}
}

\paragraph{Abbreviated Formulas.}

Formulas are usually abbreviated in a natural way.
The associativity and precedence rules for abbreviating
formulas and the mechanism for parsing formulas are presented in
Sec.~\ref{ops}.  Here are some examples.
{\small
\begin{tabbing}
m\=mmmmmmmmmmmmmmmmmmmm\=\kill
\>  Standard Usage \> \otter\ syntax (abbreviated) \\
\>  $\forall x P(x)$ \> \verb:all x P(x): \\
\>  $\forall x y \exists z (P(x,y,z)\;\vee\; Q(x,z))$ \>
           \verb:all x y exists z (P(x,y,z) | Q(x,z)): \\
\>  $\forall x (P(x)\wedge Q(x)\wedge R(x)\;\rightarrow\;S(x))$ \>
      \verb:all x (P(x) & Q(x) & R(x) -> S(x)):
\end{tabbing}
}
Note that if a formula has a string of identical quantifiers, all but the
first can be dropped.  For example, \verb:all x all y all z p(x,y,z):
can be shortened to \verb:all x y z p(x,y,z):.
In expressions involving the associative operations \verb:&: and \verb:|:,
extra parentheses can be dropped.
Moreover, a default precedence on the logic symbols allows
us to drop more parentheses:
\verb:<->: has the same precedence as
\verb:->:, and the rest in decreasing order are
\verb:->:,
\verb:|:,
\verb:&:,
\verb:-:.
Greater precedence means closer to the root of the term (i.e., larger scope).
For example, the following three strings represent the same formula.

{\small
\begin{verbatim}
    p | -q & r -> -s | t.
    (p | (-(q) & r)) -> (-(s) | t).
    ->(|(p,&(-(q),r)),|(-(s),t)).
\end{verbatim}
}
\noindent
When in doubt about how a particular string will be parsed, one can
simply add additional parentheses and/or test the string by having
\otter\ read it and then display it in pure prefix form.  The following
input file can be used to test the preceding example.

{\small
\begin{verbatim}
    assign(stats_level, 0).
    set(display_terms).
    formula_list(usable).
    p| -q&r-> -s|t.       % This formula has minimum whitespace.
    end_of_list.
\end{verbatim}
}
\noindent
In general, whitespace is required around \verb:all: and
\verb:exists: and to the left of \verb:-:; otherwise, whitespace
around the logic symbols can be removed.  See Sec.~\ref{ops}
for the rules.

\subsection{Infix, Prefix, and Postfix Expressions} \label{ops}

Many Prolog systems have a feature
that allows users to declare that particular function or predicate
symbols are infix, prefix, or postfix and to specify a precedence and
associativity so that parentheses can sometimes be dropped.
\otter\ has a similar feature.  In fact, the clause and formula
parsing routines use the feature.   Users who use only the
predeclared logic operators for clauses and formulas and the
predeclared infix equality \verb:=: can skip the rest of this section.

Prolog users who are familiar with the declaration mechanism should
note the following differences between the ordinary Prolog mechanism
and \otter's.
\begin{itemize}
\item
The predeclared operators are different.  See Table \ref{predeclared-ops}.
\item
\otter\ does not treat comma as an operator; in particular,
\verb:a,b,c: cannot be a term, as in \verb:a,b,c -> d,e,f:.
\item
\otter\ treats the quantifiers \verb:all: and \verb:exists: as
special cases, because they don't seem to fit neatly into the
standard Prolog mechanism.
\item
\otter\ requires whitespace in some cases where the Prolog systems do not.
\end{itemize}

Symbols to be treated in this special way are given a type and
a precedence.  Either \otter\ predeclares the symbol's properties,
or the user gives \otter\ a command of one of the following forms.
\begin{verse} \small
{\tt op(\em precedence\tt , \em type\tt , \em symbol\tt ).} \\
{\tt op(\em precedence\tt , \em type\tt , \em list-of-symbols\tt ).}
\end{verse}
The \verb:precedence: is an integer $i$, $0<i<1000$, and \verb:type:
is one of the following: \verb:xfx:, \verb:xfy:, \verb:yfx: (infix),
\verb:fx:, \verb:fy: (prefix), \verb:xf:, \verb:yf: (postfix).
See Table \ref{predeclared-ops} for the commands corresponding to the
predeclared symbols.

\begin{table}[htbp] \centering \small
\caption{Predeclared Symbols}  \label{predeclared-ops}
\begin{tabular}{ll}
\hline
\verb:op(800, xfy, # ).:                                  \\
\verb:op(800, xfx, ->).:   &    \verb:op(700, xfx, @<).:   \\
\verb:op(800, xfx, <->).:  &    \verb:op(700, xfx, @>).:   \\
\verb:op(790, xfy, |).:    &    \verb:op(700, xfx, @<=).:  \\
\verb:op(780, xfy, &).:    &    \verb:op(700, xfx, @>=).:  \\
                                                           \\
\verb:op(700, xfx, =).:    &    \verb:op(500, xfy, +).:    \\
\verb:op(700, xfx, !=).:   &    \verb:op(500, xfx, -).:    \\
                                                           \\
\verb:op(700, xfx, <).:    &    \verb:op(500, fx, +).:     \\
\verb:op(700, xfx, >).:    &    \verb:op(500, fx, -).:     \\
\verb:op(700, xfx, <=).:                                   \\
\verb:op(700, xfx, >=).:   &    \verb:op(400, xfy, *).:    \\
\verb:op(700, xfx, ==).:   &    \verb:op(400, xfx, /).:    \\
\verb:op(700, xfx, =/=).:  &    \verb:op(300, xfx, mod).:  \\
\hline
\end{tabular}
\end{table}

Given an expression that looks like it might be
associated in a number of ways, the relative precedence of the
operators determines, in part, how it is associated.  A symbol with higher
precedence is more dominant (closer to the root of the term), and
one with lower precedence binds more tightly.  For example, the
symbols
\verb:->:,
\verb:|:,
\verb:&:, and
\verb:-: have decreasing precedence; therefore the expression
\verb:p & - q | r -> s: is understood as \verb:((p & (-q)) | r) -> s:.

In each of the types, \verb:f: represents the
symbol, and \verb:x: and \verb:y:, which represent the expressions to
which the symbol applies, specify how terms are associated.
Given an expression involving symbols of the {\em same} precedence,
the types of the symbol determines, in part, the association.  See
Table \ref{op-types}.
\begin{table}[htbp] \centering
\caption{Symbol Types}  \label{op-types}
\begin{tabular}{lll}
\hline
\verb:xfx: & infix (binary) & don't associate \\
\verb:xfy: & infix (binary) & associate right \\
\verb:yfx: & infix (binary) & associate left \\
\verb:fx: & prefix (unary)  & don't associate \\
\verb:fy: & prefix (unary)  & associate \\
\verb:xf: & postfix (unary) & don't associate \\
\verb:yx: & postfix (unary) & associate \\
\hline
\end{tabular}
\end{table}
The following are examples of associativity:
\begin{itemize}
\item
If \verb:+: has type \verb:xfy:, then \verb:a+b+c+d: is understood as
\verb:a+(b+(c+d)):.
\item
If \verb:->: has type \verb:xfx:, then \verb:a->b->c: is not well formed.
\item
If \verb:-: has type \verb:fy:, then \verb:- - -p: is understood as
\verb:-(-(-(p))):.  (The spaces are necessary; otherwise, \verb:---: will
be parsed as single name.)
\item
If \verb:-: has type \verb:fx:, then \verb:- - -p: is not well formed.
\end{itemize}

\noindent {\it Caution:} The associativity specifications in the
infix symbol declarations say nothing about the logical associativity
of the operation, for example, whether \verb:(a+b)+c: is the same object as
as \verb:a+(b+c):.  The specifications are only about parsing
ambiguous expressions.  In most cases, when an operator is \verb:xfy:
or \verb:yfx:, it is also logically associative, but the logical
associativity is handled separately; it is built-in in the case of the
logic symbols \verb:|: and \verb:&: in \otter\ clauses and formulas,
and it must be axiomatized in other cases.

\paragraph{Details of the Symbol Declarations.}
(This paragraph can be skipped by most users.)
The precedence of symbols extends to the precedence of expressions in
the following way.  The precedence of an atomic, parenthesized, or
standard application expression is 0.  Respective examples are
\verb:p:, \verb:(x+y):, and \verb:p(a+b,c,d):.  The precedence of a
(well-formed) nonparenthesized nonatomic expression is the same
as the precedence of the root symbol.  For example, \verb:a&b: has
the precedence of \verb:&:, and \verb:a&b|c: has the precedence of the
greater symbol.  In the type specifications, \verb:x: represents an
expression of lower precedence than the symbol, and \verb:y:
represents an expression with precedence less than or equal to the
symbol.  Consider \verb:a+b+c:, where \verb:+: has type \verb:xfy:;
if association is to the left, then the second occurrence of \verb:+:
does {\em not} fit the type, because \verb:a+b:, which corresponds to
\verb:x:, does not have a lower precedence than \verb:+:; if association
is to the right, then all is well.  If we extend the example, under the
declarations \verb:op(700, xfx, =): and \verb:op(500, xfy, +):,
the expression \verb:a+b+c=d+e: must be understood as \verb:(a+ (b+c))= (d+e):.

\subsection{Whitespace in Expressions}

The reason for separating ordinary names from special names (Sec.
\ref{names}) is so that some whitespace (spaces, tabs, newline, and comments)
can be removed.  We can write \verb:a+b+c: (instead of having to write
\verb:a + b + c:), because ``\verb:a+b+c:'' cannot be a name, that is, it
must be parsed into five names.

\noindent{\em Caution.}  There is a deficiency in \otter's parser
having to do with whitespace between a name and opening parenthesis.
The rule to use is: {\em Insert some white space if and only if it is
{\em not} a standard application.}  For example, the two pieces of
white space in \verb:(a+ (b+c))= (d+e): are required, and no white space
is allowed after \verb:f: or \verb:g: in \verb:f(x,g(x)):.

\subsection{Bugs and Other Anomalies in the Input and Output of Expressions}

\begin{itemize}
\item
The symbol \verb:|: is either Prolog-style list punctuation or part of
a special name.  With the built-in declaration of \verb:|: as infix,
the term \verb:[a|b]: is ambiguous, with possible interpretations
$t_1=$\verb:$cons(a,b): and $t_2=$\verb:$cons(|(a,b),$nil):.
%$
\otter\ recognizes it as the first.  The term $t_2$
can be written \verb:[(a|b)]:.  The bug is that $t_2$ will be output
without the parentheses.  This is the only case we know in which
\otter\ cannot correctly read a term it has written.
\item
A term consisting of a unary \verb:+: or \verb:-: applied
to a nonnegative integer is always translated to a constant.
\item
Parsing large terms without parentheses, say
\verb:a1+a2+a3+...+a1000:, can be very slow if the operator
is left associative (\verb:yfx:).  If one intends to parse such terms,
one should make the operator right associative (\verb:xyf:).
\item
Quoted strings cannot contain a quotation mark of the same type.
\item
The flag \verb:check_arity: sometimes issues warnings when it
should not.
\item
Braces (\verb:{}:) can be used to group input expressions,
but \otter\ always uses ordinary parentheses on output.
\end{itemize}

\subsection{Examples of Operator Declarations}

\paragraph{Group Theory.}  Suppose we like to see group theory expressions
in the form $(ab^{-1}c^{-1-1})^{-1}$, in which right association is assumed.
We can approximate this for \otter\
with \verb:(a*b^ *c^ ^)^:.  (We have to make the group operator explicit;
\verb:-1: is not a legal \otter\ name; the whitespace shown is required.)
The declarations \verb:op(400, xfy, *): and \verb:op(350, yf, ^):
suffice.  Other examples of expressions (with minimum whitespace)
using these declarations are \verb:(x*y)*z=x*y*z: and \verb:(y*x)^ =x^ *y^:.

\paragraph{\otter\ Options.}  Options are normally input (Sec.~\ref{input-options})
as in the following examples.
\begin{verse} \small
\verb:set(prolog_style_variables):. \\
\verb:clear(print_kept):. \\
\verb:assign(max_given, 300):.
\end{verse}
If, however, we make the declarations
(the precedences are irrelevant in this case)
\begin{verse} \small
\verb:op(100, fx, set).: \\
\verb:op(100, fx, clear).: \\
\verb:op(100, xfx, assign).: \\
\end{verse}
then we may write
\begin{verse} \small
\verb:set prolog_style_variables:. \\
\verb:clear print_kept:. \\
\verb:max_given assign 300:.
\end{verse}

\section{Commands and the Input File} \label{commands}

Input to \otter\ consists of a small set of commands, some of
which indicate that a list of objects (clauses, formulas, or weight
templates) follows the command.  All lists of objects are terminated
with \verb:end_of_list:.  The commands are given in Table \ref{command-tab}.
There are a few other commands for fringe features (Sec.~\ref{fringe}).
\begin{table}[ht] \centering \small
\caption{Commands}  \label{command-tab}
\begin{tabular}{ll}
\hline
{\tt include({\it file\_name}).}   & \verb:% read input from another file:  \\
{\tt op({\it precedence}, {\it type}, {\it name(s)}).} & \verb:% declare operator(s):  \\
{\tt make\_evaluable({\it sym}, {\it eval-sym}).} & \verb:% make a symbol evaluable:  \\
{\tt set({\it flag\_name}).}           & \verb:% set a flag: \\
{\tt clear({\it flag\_name}).}         & \verb:% clear a flag: \\
{\tt assign({\it parameter\_name},{\it integer}).} & \verb:% assign to a parameter: \\
{\tt list({\it list\_name}).}          & \verb:% read a list of clauses: \\
{\tt formula\_list({\it list\_name}).} & \verb:% read a list formulas: \\
{\tt weight\_list({\it weight\_list\_name}).}  & \verb:% read weight templates: \\
{\tt lex({\it symbol\_list}).}         & \verb:% assign an ordering on symbols: \\
{\tt skolem({\it symbol\_list}).}      & \verb:% identify skolem functions: \\
{\tt lrpo\_multiset\_status({\it symbol\_list}).} & \verb:% status for LRPO: \\
\hline
\end{tabular}
\end{table}
\noindent

\subsection{Input of Options} \label{input-options}

\otter\ recognizes two kinds of option: flags and parameters.
Flags are Boolean-valued options; they are changed with the
\verb:set: and the \verb:clear: commands, which take the name of the flag as
the argument.
Parameters are integer-valued options; they are changed with the
\verb:assign: command, which takes the name of the parameter as the
first argument and an integer as the second.  Examples are

{\small
\begin{verbatim}
    set(binary_res).            % enable binary resolution
    clear(back_sub).            % do not use back subsumption
    assign(max_seconds, 300).   % stop after about 300 CPU seconds
\end{verbatim}
}
\noindent
The options are described and their default values are given in
Sec.~\ref{options}.

\subsection{Input of Lists of Clauses}

A list of clauses is specified with one of the following and
is terminated with \verb:end_of_list:.
Each clause is terminated with a period.

{\small
\begin{verbatim}
    list(usable).
    list(sos).
    list(demodulators).
    list(passive).
\end{verbatim}
}
\noindent
Example:

{\small
\begin{verbatim}
    list(usable).
      x = x.                       % reflexivity
      f(e,x) = x.                  % left identity
      f(g(x),x) = e.               % left inverse
      f(f(x,y),z) = f(x,f(y,z)).   % associativity
      f(z,x) != f(z,y) | x = y.    % left cancellation
      f(x,z) != f(y,z) | x = y.    % right cancellation
    end_of_list.
\end{verbatim}
}
\noindent
If the input contains more than one clause list of the same type,
the lists will simply be concatenated.

\subsection{Input of Lists of Formulas}

A list of formulas is specified with one of the following and
is terminated with \verb:end_of_list:.
Each formula is terminated with a period.
(Note that demodulators cannot be input as formulas.)

{\small
\begin{verbatim}
    formula_list(usable).
    formula_list(sos).
    formula_list(passive).
\end{verbatim}
}
\noindent
Example (analogous to above):

{\small
\begin{verbatim}
    formula_list(usable).
      all a (a = a).                          % reflexivity
      all a (f(e,a) = a).                     % left identity
      all a (f(g(a),a) = e).                  % left inverse
      all a b c (f(f(a,b),c) = f(a,f(b,c))).  % associativity
      all a b c (f(c,a) = f(c,b) -> a = b).   % left cancellation
      all a b c (f(a,c) = f(b,c) -> a = b).   % right cancellation
    end_of_list.
\end{verbatim}
}
\noindent
If the input contains more than one formula list of the same type,
the lists will simply be concatenated.

\subsection{Input of Lists of Weight Templates} \label{input-weight}

A list of weight templates is specified with one of the following and
is terminated with \verb:end_of_list:.
Each weight template is terminated with a period.

{\small
\begin{verbatim}
    weight_list(pick_given).      % to select given clauses
    weight_list(purge_gen).       % to discard generated clauses
    weight_list(pick_and_purge).  % to both pick and purge
    weight_list(terms).           % to order terms
\end{verbatim}
}
\noindent
Example:

{\small
\begin{verbatim}
    weight_list(pick_and_purge).
      weight(a, 0).                  % weight of constant a is 0
      weight(g($(2)), -50).          % twice weight of arg -50
      weight(P($(1),$(1)), 100).     % sum of weights of args +100
      weight(x, 5).                  % all variables have weight 5
      weight(f(g($(3)),$(4)), -300). % see Sec. ``Weighting''
    end_of_list.
\end{verbatim}
}
%$
\noindent
See Sec.~\ref{weighting} for the syntax and use of weight templates.

\subsection{The Commands {\tt lex}, {\tt skolem},
               and {\tt lrpo\_multiset\_status}} \label{symbol-commands}

Each of the commands \verb:lex:, \verb:skolem:, and \verb:lrpo_multiset_status:
takes a list of terms as an argument.  The
\verb:lex: command specifies an ordering on symbols, and the
others give properties to symbols.  An example is

{\small
\begin{verbatim}
    lex( [a, b, f(_,_), d, g(_), c] ).
\end{verbatim}
}
\noindent
The arguments of \verb:f: and \verb:g: serve as place-holders only;
they identify \verb:f: and \verb:g: as function or predicate symbols
and specify the arity.

\begin{description}

\item[{\tt lex([...])}.]
The \verb:lex: command specifies an ordering (smallest-first) on
function and constant symbols.  Lexical ordering on terms is used in
four contexts: orienting equality literals (Secs. \ref{orient} and
\ref{orient-lrpo}), deciding whether an equality will be used
as a demodulator (Secs. \ref{dynamic} and \ref{dynamic-lrpo}),
deciding whether to apply a lex-dependent
demodulator (Secs. \ref{lex-dep} and \ref{lex-dep-lrpo}), and
evaluating functions/predicates that perform lexical comparisons
(Sec.~\ref{eval}).  If a \verb:lex: command is not present, then
\otter\ uses a default ordering (Sec.~\ref{ordering}).

\item[{\tt skolem([...])}.]
The \verb:skolem: command identifies constant and function symbols as
Skolem symbols.  (If the user inputs quantified formulas and
\otter\ Skolemizes, this command is not necessary.)  The Skolem
property is used by the options \verb:para_skip_skolem:
(Sec.~\ref{para-flags}) and \verb:delete_identical_nested_skolem:
(Sec.~\ref{gen-flags}).

\item[{\tt lrpo\_multiset\_status([...])}.]
This command specifies multiset status for the lexicographic recursive
path ordering (flag \verb:lrpo:).  See Sec.~\ref{lrpo}.
\end{description}

\subsection{Other Commands}

The command {\tt op({\it precedence}, {\it type}, {\it name(s)})},
example \verb:op(400,xfy,+):, declares one or more symbols to have
special properties with respect to input and output.  See
Sec.~\ref{ops}.

\begin{sloppypar}
The command {\tt make\_evaluable({\it symbol}, {\it
evaluable-symbol})}, for example \verb:make_evaluable(_+_, $SUM(_,_):,
copies evaluation properties from an evaluable symbol to another
symbol, so that one can write \verb:x+3: instead of \verb:$SUM(x,3):.
See Sec.~\ref{make-eval}.
\end{sloppypar}

The command {\tt include({\it file\_name})} causes input to be
read from another input file.  When the included file has been
read, \otter\ resumes reading commands after the \verb:include:
command.
The file name must be recognized as an \otter\ name, so if it
contains characters such as period, slash, or hyphen, it must
be enclosed in (single or double) quotes.
Included files can include still other files.
{\em A list of objects (clauses, formulas, or weight templates)
cannot be split among different input files.}  One can, however,
read clauses into a list from more than one file, as in the
following example.

\begin{center}
\begin{tabular}{l|l|l}
standard input           & file f1.in           & file f2.in           \\
\hline
\verb:include("f1.in").: & \verb:list(usable).: & \verb:list(usable).: \\
\verb:include("f2.in").: & \verb:p(a).:         & \verb:p(b).:         \\
                         & \verb:end_of_list.:  & \verb:end_of_list.:
\end{tabular}
\end{center}

\section{Options} \label{options}

Flags are Boolean-valued options, and parameters are integer-valued
options.  When the user changes an option, \otter\ sometimes
automatically changes other options.  The user is informed in the
output file when such a change occurs.

Several additional flags and parameters are described in Sec.~\ref{fringe}.

\subsection{Flags}

Flags are changed with the \verb:set: and \verb:clear: commands, for
example,

{\small
\begin{verbatim}
    set(sos_queue).
    clear(print_given).
\end{verbatim}
}

\subsubsection{Main Loop Flags} \label{loop-flags}

A given clause is taken from \verb:sos: at the beginning of each
iteration of the main loop.  The default is to take the lightest
clause with respect to either \verb:weight_list(pick_given): or
\verb:weight_list(pick_and_purge):.  If neither weight list is present,
the weight of a clause is its number of symbols.

\noindent
\verb:sos_queue:.  Default clear.  If this flag is set,
the first clause in \verb:sos: is selected as the given clause (the set of
support list operates as a queue).  This causes a breadth-first
search, also called level saturation.  Some information about search
levels is printed (see Sec.~\ref{output}) if this flag is set.

\noindent
\verb:sos_stack:.  Default clear.  If this flag is set,
the last clause in \verb:sos: becomes the given clause (the set
of support list operates as a stack).  This causes a depth-first
search (which rarely is useful with \otter).

\noindent
\verb:input_sos_first:.  Default clear.  If this flag is set, the
input clauses in \verb:sos: are given a very low \verb:pick_given:
weight so that they are the first clauses selected as given clauses.

\noindent
\verb:interactive_given:.  Default clear.  If this flag is set,
then when it's time to select a new given clause,
the user is prompted for a choice.  This flag has priority
over all other flags that govern selection of the given clause.

\noindent
\verb:print_given:.  Default set.  If this flag is set, clauses are
output when they become given clauses.

\noindent
\verb:print_lists_at_end:.  Default clear.  If this flag is set,
then \verb:usable:, \verb:sos:, and \verb:demodulators: are printed at
the end of the search.

\subsubsection{Inference Rules} \label{inf-flags}

\verb:binary_res:.  Default clear.  If this flag is set,
the inference rule binary
resolution (along with any other inference rules that are set)
is used to generate new clauses.
Setting this flag causes the flags \verb:factor: and
\verb:unit_deletion: to be automatically set.

\noindent
\verb:hyper_res:.  Default clear.  If this flag is set,
the inference rule (positive) hyperresolution (along with any
other inference rules that are set) is used to generate new clauses.

\noindent
\verb:neg_hyper_res:.  Default clear.  If this flag is set,
the inference rule negative hyperresolution (along with any
other inference rules that are set) is used to generate new clauses.

\noindent
\verb:ur_res:.  Default clear.  If this flag is set,
the inference rule UR-resolution (unit-resulting resolution)
(along with any other inference rules that are set) is used
to generate new clauses.

\noindent
\verb:para_into:.  Default clear.  If this flag is set,
the inference rule ``paramodulation {\it into} the given clause''
(along with any other inference rules that are set) is used to
generate new clauses.  When using paramodulation, one should include
the appropriate clause for reflexivity of equality, for example, \verb:x=x:.

\noindent
\verb:para_from:.  Default clear.  If this flag is set,
the inference rule ``paramodulation {\it from} the given clause''
(along with any other inference rules that are set) is used to
generate new clauses.  When using paramodulation, one should include
the appropriate clause for reflexivity of equality, for example, \verb:x=x:.

\noindent
\verb:demod_inf:.  Default clear.
If this flag is set, demodulation is applied, as if it were an inference
rule, to the given clause.  This is useful when term rewriting
is the main objective.  When this flag is set, the given clause is copied,
then processed just like any newly generated clause.

\subsubsection{Resolution Restriction Flags} \label{res-flags}

\noindent
\verb:order_hyper:.  Default set.  If this flag is set, then
the inference rules \verb:hyper_res: and \verb:neg_hyper_res: are
constrained by an ordering strategy.  A literal in a satellite is
allowed to resolve only if it is maximal in the satellite.  (A literal
is maximal in a clause if and only if there is no larger literal.)
The ordering uses only the lexical value (as in the \verb:lex: command
or the default, Sec.~\ref{symbol-commands}) of the predicate symbol.
(This flag is irrelevant for positive hyperresolution with a Horn set.)

\noindent
\verb:unit_res:.  Default clear.  
This flag is a restriction on binary resolution.
If it is set, then all binary resolution inferences must
be unit resolutions; that is, one of the parents must be a
unit clause.  Setting this flag causes to the flag
\verb:binary_res: to be set as well.

\noindent
\verb:ur_last:.  Default clear.
This flag is a restriction on unit-resulting resolution.
If it is set, then the UR-resolvent must come from
the last literal of the nonunit parent (the nucleus).
This is related to the \emph{target strategy} in
linked UR-resolution.

\subsubsection{Paramodulation Restriction Flags} \label{para-flags}

\verb:para_from_left:.  Default set.  If this flag is set,
paramodulation is allowed {\it from} the left sides of equality literals.
(Applies to both \verb:para_into: and \verb:para_from: inference rules.)

\noindent
\verb:para_from_right:.  Default set.  If this flag is set,
paramodulation is allowed {\it from} the right sides of equality literals.
(Applies to both \verb:para_into: and \verb:para_from: inference rules.)

\noindent
\verb:para_into_left:.  Default set.  If this flag is set,
paramodulation is allowed {\it into} left sides of positive and negative
equalities.
(Applies to both \verb:para_into: and \verb:para_from: inference rules.)

\noindent
\verb:para_into_right:.  Default set.  If this flag is set,
paramodulation is allowed {\it into} right sides of positive and negative
equalities.  (Applies to both
\verb:para_into: and \verb:para_from: inference rules.)

\noindent
\verb:para_from_vars:.  Default clear.  If this flag is set,
paramodulation {\it from} variables is allowed.
{\em Warning: Setting this option may produce too many paramodulants.}
(Applies to both \verb:para_into: and \verb:para_from: inference rules.)

\noindent
\verb:para_into_vars:.  Default clear.  If this flag is set,
paramodulation {\it into} variables is allowed.
{\em Warning: Setting this option may produce too many paramodulants.}
(Applies to both \verb:para_into: and \verb:para_from: inference rules.)

\noindent
\verb:para_from_units_only:.  Default clear.  If this flag is set,
paramodulation is allowed only if the {\it from} clause is a unit (equality).
(Applies to both \verb:para_into: and \verb:para_from: inference rules.)

\noindent
\verb:para_into_units_only:.  Default clear.  If this flag is set,
paramodulation is allowed only if the {\it into} clause is a unit.
(Applies to both \verb:para_into: and \verb:para_from: inference rules.)

\noindent
\verb:para_skip_skolem:.  Default clear.  If this flag is set,
paramodulation is never allowed {\it into} subterms of Skolem
expressions \cite{skolem-aaai}.  (Applies to both \verb:para_into: and
\verb:para_from: inference rules.)

\noindent
\verb:para_ones_rule:.  Default clear.  If this flag is set,
paramodulation obeys the 1's rule.  (The 1's rule is a special-purpose
strategy for problems in combinatory logic; its usefulness has not
been demonstrated elsewhere.)  (Applies to both \verb:para_into: and
\verb:para_from: inference rules.)

\noindent
\verb:para_all:.  Default clear.  If this flag is set,
all occurrences of the {\it into} term are replaced with the
replacement term.  (Applies to both \verb:para_into: and
\verb:para_from: inference rules.)

\subsubsection{Flags for Handling Generated Clauses} \label{gen-flags}

(Section~\ref{eq-flags} describes equality-related flags for handling
generated clauses.)

\noindent
\verb:detailed_history:.  Default set.  This flag affects the
parent lists in clauses that are derived by \verb:binary_res:,
\verb:para_from:, or \verb:para_into:.  If the flag is set,
the positions of the unified literals or terms are given
along with the IDs of the parents.  See Sec.~\ref{output}
for examples.

\noindent
\verb:order_history:.  Default clear.  This flag affects the
order of parent lists in clauses that are derived by hyperresolution,
negative hyperresolution, or UR-resolution.  If the flag is set, then
the nucleus is listed first, and the satellites are listed in the order
in which the corresponding literals appear in the nucleus.  If the flag
is clear (or if the clause was derived by some other inference rule),
the given clause is listed first.

\noindent
\verb:unit_deletion:.  Default clear.
If this flag is set, unit deletion is applied to newly generated clauses.
Unit deletion removes a literal from a newly
generated clause if the literal is the negation of an instance of a
unit clause that occurs in \verb:usable: or \verb:sos:.
For example, the second literal of \verb:p(a,x) | q(a,x): is removed by
the unit \verb:-q(u,v):; but it is not removed by the unit \verb:-q(u,b):,
because that unification causes the instantiation of \verb:x:.
All such literals are removed from the newly generated clause, even if
the result is the empty clause.
One can view unit deletion with unit clause \verb:P: as demodulation
applied to literals with the demodulator \verb:P = $T:.
%$
(Unit deletion is not useful if all generated clauses are units.)

\noindent
\verb:back_unit_deletion:.  Default clear.
If this flag is set, then whenever a unit clause is derived
and kept, it is used to apply unit deletion to all existing
clauses in \verb:usable: or \verb:sos:.

\noindent
\verb:delete_identical_nested_skolem:.  Default clear.  If this
flag is set, clauses with the nested Skolem property are deleted.
A clause has the nested Skolem property if it contains a
a Skolem expression that (properly) contains
an occurrence of its leading Skolem symbol.  For example, if \verb:f:
is a Skolem function, a clause containing a term \verb:f(f(x)):
or a term \verb:f(g(f(x))): is deleted.

\noindent
\verb:sort_literals:.  Default clear.  If this flag is set,
literals of newly generated clauses are sorted---negative
literals, then positive literals, then answer literals.
The main purpose of this flag is to make clauses more readable.
In some cases, this flag can speed up subsumption on non-unit clauses.

\noindent
\verb:for_sub:.  Default set.  If this flag is set, forward
subsumption is applied during the processing of newly generated clauses.
(New clauses are deleted if subsumed by any clause
in \verb:usable: or \verb:sos:.)

\noindent
\verb:back_sub:.  Default set.  If this flag is set,
back subsumption is applied during the processing of newly kept clauses.
(Clauses in \verb:usable: or \verb:sos: are deleted if subsumed
by the newly kept clause.)

\noindent
\verb:factor:.  Default clear.  If this flag is set, factoring is
applied in two ways.
First, factoring is applied as a simplification rule
to newly generated clauses.  If a generated clause $C$ has factors
that subsume $C$, it is replaced with its smallest subsuming factor.
Second, it is applied as an inference rule to newly kept clauses.
Note that unlike other inference rules, factoring is
not applied to the given clause; it is applied to a new clause as
soon as it is kept.  All factors are generated in an iterative manner.
Factoring {\em is} attempted on answer literals.  If \verb:factor: is
set, a clause with $n$ literals will {\em not} cause a clause with fewer
than $n$ literals to be deleted by subsumption.

\subsubsection{Demodulation and Ordering Flags} \label{eq-flags}

\verb:demod_history:.  Default set.  If this flag is set, then when
a clause is demodulated, the ID numbers of the demodulators are included
in the derivation history of the clause.

\noindent
\verb:order_eq:.  Default clear.  If this flag is set,
equalities are flipped if the right side is heavier than the left.
See Secs. \ref{orient} and \ref{orient-lrpo} for the meaning of ``heavier''.

\noindent
\verb:eq_units_both_ways:.  Default clear.
If this flag is set, unit equality clauses
(both positive and negative)
are sometimes stored in both orientations; the action taken depends
on the flag \verb:order_eq:.
If \verb:order_eq: is clear, then whenever a unit, say $\alpha=\beta$,
is processed, $\beta=\alpha$ is automatically generated and processed.
If \verb:order_eq: is set, then the reversed equality is generated only if
the equality cannot be oriented (see Secs. \ref{orient} and \ref{orient-lrpo}).

\noindent
\verb:demod_linear:.  Default clear.  If this flag is set,
demodulation indexing is disabled, and a linear search of \verb:demodulators:
are used when rewriting terms.
With indexing disabled, if more than one
demodulator can be applied to rewrite a term, then the one whose clause
number is lowest is applied; this flag is useful when
demodulation is used to do ``procedural'' things.  With indexing
enabled (the default), demodulation is much faster, but the order in
which \verb:demodulators: is applied is not under the control of the
user.

\noindent
\verb:demod_out_in:.  Default clear.  If this flag is set,
terms are demodulated outside-in, left to right.  In other words, the
program attempts to rewrite a term before rewriting (left to right)
its subterms.  The algorithm is ``repeat \{rewrite the leftmost
outermost rewritable term\} until no more rewriting can be done or
the limit is reached''.  (The effect is like a standard reduction in
lambda calculus or in combinatory logic.)  If this flag is clear,
terms are demodulated inside-out (all subterms are fully demodulated
before attempting to rewrite a term).  (The evaluable conditional term
{\tt \$IF({\it condition},{\it then-value},{\it else-value})} is an
exception when inside-out demodulation is in effect.  See Sec.~\ref{eval}.)

\noindent
\verb:dynamic_demod:.  Default clear.  If this flag is set,
{\it some} newly kept equalities are made into demodulators (Secs.
\ref{dynamic} and \ref{dynamic-lrpo}).
Setting this flag automatically sets the flag \verb:order_eq:.

\noindent
\verb:dynamic_demod_all:.  Default clear.  If this flag is set,
\otter\ attempts to make {\it all} newly kept equalities into demodulators
(Sec.~\ref{dynamic}).  Setting this flag automatically sets the flags
\verb:dynamic_demod: and \verb:order_eq:.

\noindent
\verb:dynamic_demod_lex_dep:.  Default clear.  If this flag is set,
dynamic demodulators may be lex-dependent or {\sc lrpo}-dependent.
See Secs. \ref{dynamic} and \ref{dynamic-lrpo}.

\noindent
\verb:back_demod:.  Default clear.  If this flag is set, back
demodulation is applied to \verb:demodulators:, \verb:usable:, and
\verb:sos: whenever a new demodulator is added.  Back demodulation is
delayed until the inference rules are finished generating clauses from
the current given clause (delayed until \verb:post_process:).  Setting
the
\verb:back_demod: flag automatically sets the flags \verb:order_eq: and
\verb:dynamic_demod:.

\noindent
\verb:anl_eq:.  Default clear.
If this flag is set, a standard equational strategy will be
applied to the search.
This flag is really a metaflag; its only effect is
to alter other flags as follows:
{\small
\verb:set(para_from):,
\verb:set(para_into):,
\verb:set(para_from_left):,
\verb:clear(para_from_right):,
\verb:set(para_into_left):,
\verb:clear(para_into_right):,
\verb:set(para_from_vars):,
\verb:set(eq_units_both_ways):,
\verb:set(dynamic_demod_all):,
\verb:set(back_demod):,
\verb:set(process_input):, and
\verb:set(lrpo):}.
This strategy is derived mostly from equational strategies
developed at Argonne by Larry Wos and Ross Overbeek.
It can also be used for Knuth-Bendix completion.
See Sec.~\ref{kb} for more details.

\noindent
\verb:knuth_bendix:.  Default clear.  Setting this flag
simply causes the preceding flag, \verb:anl_eq:, to be set.

\noindent
\verb:lrpo:.  Default clear.  If this flag is set, then the
lexicographic recursive path ordering (also called {\sc rpo} with
status) is used to compare terms.  If this flag is clear, weight
templates and lexicographic order are used (Secs. \ref{lrpo}
and \ref{kb}).

\noindent
\verb:lex_order_vars:.  Default clear.  This flag affects
lex-dependent demodulation and the evaluable functions and predicates
that perform lexical comparisons.  If this flag is set, then lexical
ordering is a total order on terms; variables are lowest in the term
order, with \verb:x: $\prec$ \verb:y: $\prec$ \verb:z: $\prec$ \verb:u:
$\prec$ \verb:v: $\prec$ \verb:w: $\prec$ \verb:v6: $\prec$ \verb:v7:
$\prec$ \verb:v8: $\prec$ $\cdots$.  If this flag is clear, then a
variable is comparable only to another occurrence of the same
variable; it is not comparable to other variables or to nonvariables.
For example, \verb:$LLT(f(x),f(y)): evaluates to \verb:$T: if and only
if \verb:lex_order_vars: is set.  {\it If \verb:lrpo: is set,
\verb:lex_order_vars: has no effect on demodulation}  (Sec.
\ref{lex-order}).

\noindent
\verb:symbol_elim:.  Default clear.  If this flag is set, then
new demodulators are oriented, if possible, so that function symbols
(excluding constants) are eliminated.  A demodulator can eliminate all
occurrences of a function symbol if the arguments on the left side are
all different variables and if the function symbol of the left side does
not occur in the right side.  For example, the demodulators
\verb:g(x) = f(x,x): and \verb:h(x,y) = f(x,f(y,f(g(x),g(y)))): eliminate all
occurrences of \verb:g: and \verb:h:, respectively.

\noindent
\verb:rewriter:.  Default clear.  If this flag is set, then the clauses
in the \verb:sos: list will simply be demodulated by the demodulators,
the run will terminate.  This is really just a metaflag, which
automatically causes the several other options parameters to be changed
as follows:
\verb:set(demod_inf):,
\verb:clear(for_sub):,
\verb:clear(back_sub):, and
\verb:assign(max_levels, 1):.

\subsubsection{Input Flags}

\verb:check_arity:.  Default set.  If this flag is set, 
a warning is given if symbols
have variable arities (different numbers of arguments in different
places in the input).  For example, the term \verb:f(a,a(b)): would
be flagged.  (Constants have arity 0.)  If this flag is clear, then
variable arities are permitted; in the preceding term, the two
occurrences of \verb:a: would be treated as different symbols.

\noindent
\verb:prolog_style_variables:.  Default clear.  If this flag is
set, a name with no arguments in a clause is a variable if and only if
it starts with \verb:A: through \verb:Z: (upper case) or with \verb:_:.

\noindent
\verb:echo_included_files:.  Default set.  If this flag is
set, input files included with the {\tt include(\it filename\tt )}
command are echoed in the same way as ordinary input.  

\noindent
\verb:simplify_fol:.  Default set.  If this flag is set, then
some propositional simplification is attempted when converting input
first-order formulas into clauses.  The simplification occurs after
Skolemization, during the CNF translation.  If simplification
detects a refutation, it will always produce the empty clause \verb:$F:,
%$
but \otter\ will not recognize the proof (i.e., give the proof
message and stop) unless the flag \verb:process_input: is set.

\noindent
\verb:process_input:.  Default clear.  If this flag is set, input
\verb:usable: and \verb:sos: clauses (including clauses from formula
input) are processed as if they had been generated by an inference rule.
(See the procedure for processing newly
inferred clauses in Sec.~\ref{outline}.)  The exceptions are
(1) the following clause-processing options are not applied
to input clauses: \verb:max_literals:, \verb:max_weight:,
\verb:delete_identical_nested_skolem:, and \verb:max_distinct_vars:,
(2) clauses input on list {\tt usable} remain there if retained, and
(3) some output appears even if the output flags
(Sec.~\ref{output-flags}) are clear.

\noindent
\verb:tptp_eq:.  Default clear.  If this flag is set, then
``\verb:EQUAL:'' is the one and only symbol recognized as
the equality relation for the operations that build in equality
(demodulation and paramodulation).

\subsubsection{Output Flags} \label{output-flags}

\noindent
\verb:very_verbose:.  Default clear.  If this flag is set,
a tremendous amount of information about the processing of generated
clauses is output.

\noindent
\verb:print_kept:.  Default set.  If this flag is set, new
clauses are output if they are retained (if they pass all retention tests).

\noindent
\verb:print_proofs:.  Default set.
If this flag is set, all proofs that are found are printed to the output file.
If this flag is clear, no proofs are printed.

\noindent
\verb:build_proof_object_1:.  Default clear.  If this flag is set,
then whenever a proof is found, a
\emph{type 1 proof object} is printed to the output file.
Proof objects are very detailed proof and were introduced for two purposes: 
so that proofs can be checked by an independent program, and
so that proofs can be translated into other forms by other programs.
Proof objects are written in a Lisp-like notation.
(Type 2 proof objects are usually preferred.)
\emph{Warning: Construction of proof objects is fragile---sometimes it
simply fails.}

\noindent
\verb:build_proof_object_2:.  Default clear.  If this flag is set,
then whenever a proof is found, a
\emph{type 2 proof object} is printed to the output file.
Type 2 proof objects are used in the IVY verification project
\cite{ivy}, and a detailed description (definition in ACL2)
can be found there.
\emph{Warning: construction of proof objects is fragile---sometimes it
simply fails.}

\noindent
\verb:print_new_demod:.  Default set.  If this flag is set,
demodulators that are adjoined during the search (\verb:dynamic_demod:)
are printed.
New demodulators are always printed during input processing.

\noindent
\verb:print_back_demod:.  Default set.  If this flag is set,
clauses are printed as they are back demodulated.
Back-demodulated clauses are always printed during input processing.

\noindent
\verb:print_back_sub:.  Default set.  If this flag is set,
clauses are printed if they are back subsumed.
Back-subsumed clauses are always printed during input processing.

\noindent
\verb:display_terms:.  Default clear.  If this flag is set, all
clauses and terms are printed in pure prefix form (Sec.~\ref{syntax-terms}).
This feature can be useful for debugging the input.

\noindent
\verb:pretty_print:.  Default clear.  If this flag is set, clauses
are output in an indented form that is sometimes easier to read.
The parameter \verb:pretty_print_indent: (default 4) specifies the
number of spaces for each indent level.

\noindent
\verb:bird_print:.  Default clear.  If this flag is set,
terms constructed with the binary function \verb:a: are output in combinatory
logic notation (without the function symbol \verb:a:, and left
associated unless otherwise indicated).  For example, the clause
\verb:a(a(a(S,x),y),z) = a(a(x,z),a(y,z)): is output as
\verb:S x y z = x z (y z):.
Terms cannot be input in combinatory logic notation.

\noindent
\verb:formula_history:.  Default clear.
If this flag is set, and if quantified formulas are given as input,
then the formulas will occur in proofs, and the clauses derived
from the formulas will refer to the formulas with the justification
\verb:clausify:.

\subsubsection{Indexing Flags} \label{index-flags}

\verb:index_for_back_demod:.  Default set.  If this flag is set,
all nonvariable terms in all clauses are indexed so that
the appropriate ones can be quickly retrieved when applying a
dynamic demodulator to the clause space (back demodulation).  This
type of indexing can use a lot of memory.  If the flag is clear,
back demodulation still works, but it is much slower.

\noindent
\verb:for_sub_fpa:.  Default clear.  If this flag is set,
{\sc fpa} indexing is used for forward subsumption.  If this flag is
clear, discrimination tree indexing is used.  Setting this
flag can decrease the amount of memory required by \otter.
Discrimination tree indexing can require a lot of memory, but it is
usually {\em much} faster than {\sc fpa} indexing.

\noindent
\verb:no_fapl:.  Default clear.  If this flag is set, 
positive literals are not indexed for unit conflict or back
subsumption.  This option should be used only when no negative units will be
generated (as with hyperresolution), back subsumption is disabled, and
discrimination tree indexing is being used for forward subsumption.
This option can save a little time and memory.

\noindent
\verb:no_fanl:.  Default clear.  If this flag is set,
negative literals are not indexed for unit conflict or back
subsumption.  This option should be used only when no positive units
will be generated (as with negative hyperresolution), back subsumption
is disabled, and discrimination tree indexing is being used for
forward subsumption.  This option can save a little time and memory.

\subsubsection{Miscellaneous Flags} \label{misc-flags}

\verb:control_memory:.  Default clear.  If this flag is set, then
the automatic memory-control feature is enabled (Sec.~\ref{mem-control}).

\noindent
\verb:propositional:.  Default clear.  If this flag is
set, \otter\ assumes that all clauses are propositional,
and it makes some optimizations.
{\em The user should set this flag only when all clauses are
propositional; otherwise \otter\ may make unsound inferences and/or
crash.}

\noindent
\verb:really_delete_clauses:.  Default clear.  If this flag is
clear, clauses that are deleted by back subsumption or back
demodulation are not really removed from memory; they are retained in
a special place so that they can be printed if they occur in a proof.
If the job involves much back subsumption or back demodulation and if
memory conservation is important, these ``deleted'' clauses can be removed
from memory by setting this flag (and any proof containing such a
clause will not be printed in full).

\noindent
\verb:atom_wt_max_args:.  Default clear.  If this flag is set, the
default weight of an atom (the weight if no template matches the atom)
is 1 plus the maximum of the weights of the arguments.  If this flag is
clear, the default weight of an atom is 1 plus the sum of the weights of
the arguments.

\noindent
\verb:term_wt_max_args:.  Default clear.  If this flag is set, the
default weight of a term (the weight if no template matches the atom)
is 1 plus the maximum of the weights of the arguments.  If this flag is
clear, the default weight of a term is 1 plus the sum of the weights of
the arguments.

\noindent
\verb:free_all_mem:.  Default clear.  If this flag is set, then
at the end of the search, most dynamically allocated memory is
returned to the memory managers.  This flag is used mainly for
debugging, in particular, to help find memory leaks.
Setting this flag will {\em not} cause \otter\ to use less memory.

\noindent
\verb:sigint_interact:.  Default set.  If this flag is set, then
when \otter\ receives an interrupt signal from the operating
system (usually caused by the user pressing control-C),
\otter\ will enter a primitive interactive mode, which is described in
Sec.~\ref{interact}.
\newpage
\subsection{Parameters}

Parameters are integer-valued options.
In the descriptions that follow,
\maxint\ is a large integer, usually the size of the largest
ordinary integer on the user's computer (i.e., INT\_MAX in ANSI C).

\subsubsection{Monitoring Progress}

\verb:assign(report,:$n$\verb:):.  Default $-1$, range [$-1$..\maxint ].  If $n>0$,
then statistics are output approximately every $n$ {\sc cpu} seconds.  The
time is not exact because statistics will be output only after the
current given clause is finished.  This feature can be used in
conjunction with {\sc unix} programs such as \verb:grep: and
\verb:awk: to conveniently monitor \otter\ jobs.

\subsubsection{Placing Limits on the Search}

\verb:assign(max_seconds,:$n$\verb:):.  Default $-1$, range [$-1$..\maxint ].  If $n\neq -1$,
the search is terminated after about $n$ {\sc cpu} seconds.
The time is not exact because \otter\ will wait until the current
given clause is finished before stopping.

\noindent
\verb:assign(max_gen,:$n$\verb:):.  Default $-1$, range [$-1$..\maxint ].  If $n\neq -1$,
the search is terminated after about $n$ clauses have been
generated.  The number is not exact because \otter\ will wait until it is
finished with the current given clause before stopping.

\noindent
\verb:assign(max_kept,:$n$\verb:):.  Default $-1$, range [$-1$..\maxint ].  If $n\neq -1$,
the search is terminated after about $n$ clauses have been
kept.  The number is not exact because \otter\ will wait until it is
finished with the current given clause before stopping.

\noindent
\verb:assign(max_given,:$n$\verb:):.  Default $-1$, range [$-1$..\maxint ].  If $n\neq -1$,
the search is terminated after $n$ given clauses have been
used.

\noindent
\verb:assign(max_levels,:$n$\verb:):.  Default $-1$, range [$-1$..\maxint ].  If $n\neq -1$,
the flag \verb:sos_queue: will be automatically set, causing a level saturation
(breadth-first) search.  In this case the search is terminated after $n$ levels
have been processed.

\noindent
\verb:assign(max_mem,:$n$\verb:):.  Default $-1$, range [$-1$..\maxint ].  If $n\neq -1$,
\otter\ will terminate the search before more than $n$
kilobytes have been dynamically allocated (\verb:malloc:).

\subsubsection{Limits on Properties of Generated Clauses}

\verb:assign(max_literals,:$n$\verb:):.  Default $-1$, range [$-1$..\maxint ].  If $n\neq -1$,
new clauses are discarded if they contain more than $n$
literals.

\noindent
\verb:assign(max_weight,:$n$\verb:):.  Default \maxint, range [$-$\maxint ..\maxint ].
New clauses are discarded if their weight is more than
$n$.  The weight list \verb:purge_gen: or the weight list
\verb:pick_and_purge: is used to weigh clauses (both lists may not be
present; see Sec.~\ref{weighting}).

\noindent
\verb:assign(max_distinct_vars,:$n$\verb:):.  Default $-1$, range [$-1$..\maxint ].
If $n\neq -1$,
new clauses are discarded if they contain more than $n$
distinct variables.

\noindent
\verb:assign(max_answers,:$n$\verb:):.  Default $-1$, range [$-1$..\maxint ].
If $n\neq -1$,
new clauses are discarded if they contain more than $n$
answer literals.

\subsubsection{Indexing Parameters}

\verb:assign(fpa_literals,:$n$\verb:):.  Default 8, range [0..100].  $n$ is the {\sc fpa}
indexing depth for literals.  ({\sc fpa} literal indexing is used for
resolution inference rules, back subsumption, and unit conflict.  It
is also used for forward subsumption if the flag \verb:for_sub_fpa:
is set.)  If $n=0$, indexing is by predicate symbol only; if $n=1$,
indexing looks at the predicate symbol and the leading symbols of the
arguments of the literal, and so on.  Greater indexing depth requires
more memory, but it can be faster.  Changing this parameter will not
change the clauses that are generated or kept.

\noindent
\verb:assign(fpa_terms,:$n$\verb:):.  Default 8, range [0..100].  $n$ is the {\sc fpa}
indexing depth for terms.  ({\sc fpa} term indexing is used for
paramodulation inference rules and back demodulation.)  If $n=0$,
indexing is by symbol only; if $n=1$, indexing looks at the
symbol and the leading symbols of the arguments of the term; and
so on.  Greater indexing depth requires more memory, but it can be
faster.  Changing this parameter will not change the clauses that
are generated or kept.

\subsubsection{Miscellaneous Parameters} \label{misc-parms}

\verb:assign(pick_given_ratio,:$n$\verb:):.  Default $-1$, range [$-1$..\maxint ].
This parameter causes some given clauses to be selected by weight and
others in a breadth-first manner (by age).
If $n \neq -1$, 
$n$ given clauses are are selected by (smallest
\verb:pick_given:) weight,
then the first clause in \verb:sos: is selected as given clause,
then $n$ given clauses are selected by weight, and so forth.
This
method allows heavy clauses to enter into the search while focusing
mainly on light clauses.  It combines breadth-first search
and best-first search (default selection by weight).
If $n$ is $-1$, then the clause with smallest \verb:pick_given: weight is
always selected.

\noindent
\verb:assign(age_factor,:$n$\verb:):.  Default 0, range [$-$\maxint ..\maxint ].
If $n\neq$0, then the pick-given weight of clauses is adjusted as follows.
If $g$ is the number of clauses that have been given
\emph{at the time the clause is kept},
and $n$ is the age factor, then $g / n$
(with integer division) is added to the pick-given weight of the clause.

\noindent
\verb:assign(distinct_vars_factor,:$n$\verb:):.  Default 0, range [$-$\maxint ..\maxint ].
If $n\neq$0, then the pick-given weight of clauses is adjusted as follows.
If $v$ is the number of variable in the clause,
and $n$ is the age factor, then $v / n$
(with integer division) is added to the pick-given weight of the clause.

\noindent
\verb:assign(interrupt_given,:$n$\verb:):.  Default $-1$, range [$-1$..\maxint ].  If $n>0$,
then after $n$ given clauses have been used, \otter\ goes
into its interactive mode (Sec.~\ref{interact}).

\noindent
\verb:assign(demod_limit,:$n$\verb:):.  Default 1000, range [$-1$..\maxint ].  If $n\neq -1$,
$n$ is the maximum number of rewrites that will be
applied when demodulating a clause.  The count includes \verb:$:
%$
symbol evaluation. If $n$ is $-1$, there is no limit.  A warning message
is printed if \otter\ attempts to exceed the limit.

\noindent
\verb:assign(max_proofs,:$n$\verb:):.  Default 1, range [$-1$..\maxint ].  If $n = 1$,
\otter\ will stop if it finds a proof.  If $n > 1$, then
\otter\ will not stop when it has found the first proof; instead, it
will try to keep searching until it has found $n$ proofs.  (Some of
the proofs may in fact be identical.)  (Because forward
subsumption occurs before unit conflict, a clause representing a truly
different proof may be discarded by forward subsumption before unit
conflict detects the proof.)  If $n = -1$, \otter\ will find as
many proofs as it can (within other constraints).

\noindent
\verb:assign(min_bit_width,:$n$\verb:):.  Default {\em bits-per-long}, range [0..{\em bits-per-long}].
When the evaluable bit operations (Sec.~\ref{eval}) produce a new bit
string, leading zeros are suppressed under the constraint that
$n$ is the minimum string length.
(The value \emph{bits-per-long} is the number of bits in
the C data type long integer.)

\noindent
\verb:assign(neg_weight,:$n$\verb:):.  Default 0, range [$-$\maxint ..\maxint ].
The value
$n$ is the additional weight (positive or negative) that is given to
negated literals.  Weight templates cannot be used for this purpose
because the negation sign on a literal cannot occur in weight
templates.  (Atoms, not literals, are weighed with weight templates;
see Sec.~\ref{weighting}.)

\noindent
\verb:assign(pretty_print_indent,:$n$\verb:):.  Default 4, range [0..16].
See flag \verb:pretty_print:, Sec.~\ref{output-flags}.

\noindent
\verb:assign(stats_level,:$n$\verb:):.  Default 2, range [0..4].
This indicates the level of detail of statistics printed in reports and at
the end of the search.
If $n = 0$, no statistics are output;
if $n = 1$, a few important search and time statistics are output;
if $n = 2$, all search and time statistics are output;
if $n = 3$, search, time, and memory statistics are output; and
if $n = 4$, search, time, and memory statistics and option values are output.
This parameter does not affect the speed of \otter, because all
statistics are always kept.

\noindent
\verb:assign(dynamic_demod_depth,:$n$\verb:):.  Default -1, range [$-$1 ..\maxint ].\\
\verb:assign(dynamic_demod_rhs,:$n$\verb:):.  Default  1, range [$-$\maxint ..\maxint ].\\
These two parameters work together,
allowing an extension of the ad hoc ordering when
deciding whether a new equality should be a demodulator.  (It is
not used if flag \verb:lrpo: is set.)  The equality, say $\alpha=\beta$,
is first oriented as described in Sec.~\ref{ad-hoc}.
If $wt(\beta)\leq$
\verb:dynamic_demod_rhs: and if $wt(\alpha) - wt(\beta) \geq$
\verb:dynamic_demod_depth:, then the equality can be a demodulator.
With the default values for these parameters, the behavior
is as described in Sec.~\ref{ad-hoc}

\noindent
\verb:assign(new_symbol_lex_position,:$n$\verb:):.  Default \maxint, range [1 ..\maxint ].
New symbols can be created during the search, usually by \$-evaluation.
%$
With this parameter, the user can specify where
they will occur in the symbol ordering.  If there is a \verb:lex: command,
all new symbols will have a lexical values between the $n$th and
$(n+1)$th symbol in the \verb:lex: command.  The ordering among the new
symbols is the default ordering.  This also applies to input
symbols not occurring in the \verb:lex: command.

\section{Demodulation}

Basic demodulation is straightforward, but there are many variations
and enhancements whose descriptions are scattered throughout this
manual.  This section (which is mostly redundant) lists some overall
comments on demodulation and points the reader to the appropriate
sections on variations and enhancements.

\paragraph{The Equality Symbol.}
The binary symbol \verb:=: (which can be used as an infix symbol) and
any name that starts with \verb:eq:, \verb:EQ:, or \verb:Eq:, when
used as a binary predicate symbol, is recognized as an equality
predicate by demodulation.  \emph{An exception:} if the flag
\verb:tptp_eq: is set, then \verb:EQUAL: is the one and only
equality symbol; this flag was introduced for compatibility
with the TPTP problem library \cite{tptp-web}.

\paragraph{When and How It Is Applied.}
Demodulation is applied, using equalities in the list \verb:demodulators:,
to every clause that is generated by an inference rule.
Also, when the flag \verb:demod_inf: (Sec.~\ref{inf-flags})
is set, demodulation is, in effect, treated as an inference rule.

\paragraph{Demodulation of Atomic Formulas.}  Atomic formulas (literals with any
negation sign removed) can be demodulated.  Useful examples are

{\small
\begin{verbatim}
    (x*y = x*z) = (y = z).  % one form of cancellation
    D(x,y) = D(y,x).        % lex-dependent atom demodulator
    P(junk) = $T.           % trick to get rid of a literal
\end{verbatim}
}
%$
\noindent
The appropriate clause simplification occurs if the right side of an
atom demodulator is one of the Boolean constants \verb:$T: or
\verb:$F:.  Negated literals cannot be demodulated, but the atom of a
negative literal can be demodulated.

\paragraph{Inside-out or Outside-in.}
The user has the option of having terms rewritten inside-out
or outside-in.  (See the description of the flag \verb:demod_out_in:
in Sec.~\ref{eq-flags}.)  Although the choice makes little
difference for many applications, we nearly always recommend inside-out.
Outside-in can be much faster in cases where the left side of the
demodulator has a variable not in the right side.

\paragraph{Order of Demodulators.}
By default, demodulation uses an indexing mechanism to find
demodulators that can rewrite a given term;  if more than one 
demodulator can apply, the user has no control over which
one is used.  To order the set of demodulators
for application, the user can set the flag \verb:demod_linear:
(Sec.~\ref{eq-flags}).

\paragraph{Dynamic Demodulation and Back Demodulation.}
Positive equality units derived during the search can be made into
demodulators (Secs. \ref{eq-flags}, \ref{dynamic}, and
\ref{dynamic-lrpo}).  Demodulators adjoined during the search can be
used to rewrite previously derived clauses (Sec.~\ref{eq-flags}).

\paragraph{Termination.}
With the default ad hoc ordering,
demodulation is not guaranteed to terminate by itself.  Therefore,
a parameter (\verb:demod_limit:) specifies the maximum
number of rewrite steps that will be applied to a clause.
With the lexicographic recursive path ordering (flag \verb:lrpo:),
demodulation will always terminate by itself.  (Even with \verb:lrpo:,
the parameter \verb:demod_limit: has effect because demodulation
sequences can have an unreasonable number of steps.)

\paragraph{Introduction of New Variables.}
A demodulator introduces
new variables if it has variables on the right side that do not
occur on the left.  The {\sc lrpo} flag does not allow demodulators
to introduce new variables.  The default ordering allows variable
introductions only for input demodulators.

\paragraph{Lex- and {\sc lrpo}-dependent Demodulation.}
Ordinary demodulators are used unconditionally; they usually
simplify or canonicalize regardless of the context in which
they are applied.  But some equalities that are not normally
thought of as rewrite rules can be used as such and are
applied only if the application produces a ``better'' term.
These are called lex- or {\sc lrpo}-dependent demodulators
(depending on whether the flag \verb:lrpo: is set).
For example, commutativity of an operation, say $x+y=y+x$, can
be used to rewrite $b+a$ to $a+b$ if $a+b\prec b+a$.
See Secs. \ref{eq-flags}, \ref{lex-dep}, and \ref{lex-dep-lrpo}.
Do not confuse this type of demodulation with conditional demodulation.

\paragraph{Demodulation of Evaluable Terms.}
\otter\ has many built-in function and predicate
symbols for doing arithmetic, logic operations, bit operations, and
other operations.  The evaluation of terms containing these built-in
symbols is done as a part of demodulation (Sec.~\ref{eval}).

\paragraph{Conditional Demodulation.}
Demodulators can be written with conditions as
\begin{verse}
{\em condition} \verb:->: $\alpha=\beta$.
\end{verse}
The demodulator is applied only if the condition, instantiated with
the matching substitution, demodulates to \verb:$T: (meaning {\em true}).
%$
This is a ``fringe feature'', and it has not been heavily used
(Sec.~\ref{cond-demod}).

\paragraph{Demodulation as Equational Programming.}
\otter's demodulation, especially with the evaluable symbols,
can be used as a general-purpose (although not particularly efficient
or convenient) equational programming system (Sec.~\ref{eval}).
We have not seen many cases
where this is useful in the context of a traditional refutation
search, but it has proved to be very useful for various symbolic
programming tasks, particularly with hyperresolution.

\paragraph{Demodulation to Delete Clauses.}
Demodulation can
be used as a trick to overcome one of the deficiencies of the
weighting mechanism (Sec.~\ref{weighting}) to discard undesired
clauses.  Weighting does not implement a true match (one-way
unification) operation.  If the user wishes to discard every clause
that contains an instance of a particular term, say \verb:f(x,x):, a
demodulator, say \verb:f(x,x) = junk:, can be input along with a weight
template that gives \verb:junk: a \verb:purge_gen: weight higher than
\verb:max_weight:.  (When using this and similar tricks,
the user must make sure that the clauses containing \verb:junk: are really
discarded by weighting or another means; on occasion we have
found proofs that are incorrect because they depend on \verb:junk:.)
\newpage
\section{Ordering and Dynamic Demodulation} \label{ordering}

\begin{sloppypar}
This section contains a more complete explanation of the options
\verb:lex_order_vars:, \verb:order_eq:, \verb:symbol_elim:,
\verb:dynamic_demod:, \verb:dynamic_demod_all:, \verb:lrpo:, and
\verb:dynamic_demod_lex_dep:.  It gives all the rules---built in and
optional---for orienting equality literals and deciding which
equalities will be dynamic demodulators.  \otter\ uses two kinds
of term ordering.
\begin{description}
\item[{\it ad hoc ordering.}]  This is a collection of ordering
methods that we have accumulated through many years of experimentation.
The methods do not have a substantial theoretical foundation, but they are
useful in many cases.  This is the default ordering; it is
presented in Sec.~\ref{ad-hoc}.
\item[{\sc lrpo}.]  This is the {\em lexicographic recursive path
ordering} (also called {\sc rpo} with status).  It has nice theoretical
properties and is easier to use than the ad hoc ordering, but it is
more computationally expensive.  The {\sc lrpo} ordering is enabled
with the flag \verb:lrpo:;  it is described in Sec.~\ref{lrpo}.
\end{description}
\end{sloppypar}

Both kinds of term ordering use an ordering on constant and function
symbols.  The \verb:lex: command (Sec.~\ref{symbol-commands}) is used
to assign an ordering on symbols.  For example, the command

{\small
\begin{verbatim}
    lex( [a, b, c, d, or(_,_)] ).
\end{verbatim}
}
\noindent
specifies {\tt a $\prec$ b $\prec$ c $\prec$ d $\prec$ or} ({\tt or}
is a binary symbol).  If a \verb:lex: command is given,
all constant and function symbols in terms that will be compared must
be included.  If a \verb:lex: command is not given, \otter\ uses
the following default ordering.
\begin{verse}
{\tt [\em constants\tt , \em high-arity\tt, $\cdots$, \em binary\tt , \em unary\tt]}
\end{verse}
Within arity, the lexicographic {\sc ascii} ordering (i.e., the
C library routine \verb:strcomp():) is used.

The methods for orienting equalities and for determining
dynamic and lex-dependent demodulators apply to all inferred clauses;
if the flag \verb:process_input: is set, they also apply to input
\verb:usable: and \verb:sos: clauses.

In this section, $\alpha$ and $\beta$ always refer to
the left and right arguments, respectively, of the equality literal
under consideration; $wt(\gamma)$ refers to the weight of
$\gamma$ using \verb:weight_list_terms:; $vars(\gamma)$ is the
set of variables in $\gamma$.  The symbols $\succ$ and $\prec$ are
used for several orderings; the one referred to should be clear
from the context.

Table \ref{order-tab} is a quick reference guide to the ordering mechanisms
presented in Secs. \ref{ad-hoc} and \ref{lrpo}.

{\small
\begin{table}[htbp] \centering \small
\caption{Quick Reference to Ordering}  \label{order-tab}
\begin{tabular}{|l|l||l||l|} \hline
\multicolumn{2}{|c||}{Situation}             & \multicolumn{1}{c||}{Ad Hoc} 
							      & \multicolumn{1}{c|}{LRPO} \\ \hline\hline
Input demods       & flip?                  & no              & if $\alpha\prec\beta$      \\ \cline{2-4}
                   & lex-dependent?         & if ident-x-vars & if neither is greater   \\ \hline
\multicolumn{2}{|l||}{Orienting eqs ({\tt order\_eq} set)}
					    & \parbox{1.5in}{flip if sym-elim,\\ occurs-in, or wt-lex-ord} 
                                                              & flip if $\alpha\prec\beta$ \\ \hline

                   & \verb:d_d_all: clear   & \parbox{1.5in}{if oriented, var-subset,\\ and $wt(\beta) \leq 1$}
							      & if $\alpha\succ\beta$        \\ \cline{2-4}
Dynamic demod?     & \verb:d_d_all: set     & \parbox{1.5in}{if oriented and var-subset}
							      & if $\alpha\succ\beta$        \\ \cline{2-4}
                   & lex-dependent?         & \parbox{1.5in}{if ident-x-vars and \\ {\tt dynamic\_demod\_all} set}
							      & \parbox{1.25in}{if neither is greater,\\and var-subset} \\ \hline
\multicolumn{2}{|l||}{Apply lex-dependent demod?}
                                            & lex-order($\alpha\sigma,\beta\sigma$) 
                                                              & $\alpha\sigma\succ\beta\sigma$ \\ \hline 
\multicolumn{2}{|l||}{Lex {\tt \$} evaluation}& lex-order      & lex-order                    \\ \hline
\end{tabular}
\end{table}
}

\subsection{Ad Hoc Ordering} \label{ad-hoc}

\subsubsection{Term Ordering (Ad Hoc)} \label{lex-order}

Two types of ad hoc term ordering are used: lex-order and weight-lex-order.
The user does not have a choice between these two; the one that is
applied depends on the context, as described in the following subsections.
\begin{description}
\item[{\it lex-order.}]
This is a basic lexicographic extension of the symbol order.  To compare
two terms, one reads them left to right, stopping at the
first symbols where they differ;  the relationship of those symbols
determines the term order.
The treatment of variables depends on 
the flag \verb:lex_order_vars::
\begin{description}
\item[{\tt lex\_order\_vars} is set.]
Variables are the lowest in the symbol ordering, with \verb:x: $\prec$
\verb:y: $\prec$ \verb:z: $\prec$ \verb:u: $\prec$ \verb:v: $\prec$
\verb:w: $\prec$ \verb:v6: $\prec$ \verb:v7: $\prec$ \verb:v8: $\prec$
$\cdots$.  Since the order on symbols is total (any two symbols are
comparable), the lexical order on terms is total (any two terms are
comparable).  Note that applying a substitution to a pair of terms may
change their relative order.
\item[{\tt lex\_order\_vars} is clear (the default).]
A variable is comparable only to itself and to a term that
contains the variable.
The order on terms is partial.  Note that
if $t_1 \prec t_2$, and if $\sigma$ is any substitution, then
$t_1\sigma \prec t_2\sigma$.
\end{description}
\begin{sloppypar}
\item[{\it weight-lex-order.}]
In comparing two terms, they are first weighed with \verb:weight_list_terms:.
If one term is heavier, it is greater in the order.
If the terms have equal weight, they are compared with respect to
the lex-order as if \verb:lex_order_vars: is clear.
\end{sloppypar}
\end{description}

\subsubsection{Orienting Equalities (Ad Hoc)} \label{orient}

If the flag \verb:order_eq: is set and \verb:lrpo: is clear, then
equality literals (both positive and negative) in inferred
clauses are processed as follows.  
\begin{enumerate}
\item
If the \verb:symbol_elim: flag is set and if the equality is a
symbol-eliminating type (Sec.~\ref{eq-flags}), the
equality is oriented in the appropriate direction.
\item
If one argument is a proper subterm of the other argument,
the equality is oriented so that the subterm is the right-hand argument.
\item
If one argument is greater in the weight-lex-order, say
$\gamma\succ\delta$,
the equality is oriented with $\gamma$ as the left side.
\end{enumerate}
The preceding steps do not apply to equalities input on the list
\verb:demodulators:.

\subsubsection{Determining Dynamic Demodulators (Ad Hoc)}  \label{dynamic}

A dynamic demodulator is a demodulator that is inferred rather than
input.  If either of the flags \verb:dynamic_demod: or
\verb:dynamic_demod_all: is set, the flag \verb:order_eq: will also be
set, and \otter\ will attempt to make some or all inferred positive
equality units into demodulators.  If the flag \verb:process_input:
is set, the procedure applies to input \verb:usable: and \verb:sos:
equalities.
The procedure assumes that equalities have already been oriented.
\begin{enumerate}
\item
If the flag \verb:symbol_elim: is set and if $\alpha=\beta$ is
symbol-eliminating, the equality becomes a demodulator.
\item
If $\beta$ is a proper subterm of $\alpha$, the equality becomes a
demodulator.
\item
If $\alpha\succ\beta$ in the weight-lex-order, and if 
$vars(\alpha) \supseteq vars(\beta)$,
\begin{itemize}
\item[(a)]
if \verb:dynamic_demod_all: is set, the equality becomes a
demodulator;
\item[(b)]
if \verb:dynamic_demod_all: is clear and if $wt(\beta) \leq 1$, the
equality becomes a demodulator.
\end{itemize}
\item
If \verb:dynamic_demod_lex_dep: and \verb:dynamic_demod_all: are both set,
if $\alpha$ and $\beta$ are identical-except-variables (Sec.~\ref{lex-dep}),
and if $vars(\alpha) \supseteq vars(\beta)$,
the equality becomes a lex-dependent demodulator.
\end{enumerate}

\subsubsection{Lex-dependent Demodulation (Ad Hoc)} \label{lex-dep}

Two terms are {\it identical-except-variables} if they are identical
after replacing all occurrences of variables with \verb:x:.
An input or dynamic demodulator is lex-dependent only if
$\alpha$ and $\beta$ are identical-except-variables.
(See Sec.~\ref{dynamic} for determining lex-dependent dynamic demodulators.)
A lex-dependent demodulator applies to a term only if
the replacement term is smaller in the lex-order.
In particular, \otter\ will apply a lex-dependent demodulator
$\alpha=\beta$ if and only if $\alpha\sigma\succ\beta\sigma$ in the
lex-order, where $\sigma$ is the matching substitution.

For example, in the presence of the \verb:lex: command and the
(lex-dependent) demodulators

{\small
\begin{verbatim}
    lex([a, b, c, d, or(_,_)]).

    list(demodulators).
      or(x,y) = or(y,x).
      or(x,or(y,z)) = or(y,or(x,z)).
    end_of_list.
\end{verbatim}
}
\noindent
the term \verb:or(or(d,b),or(a,c)): will be demodulated to
\verb:or(a,or(b,or(c,d))): (in several steps).

\subsection{LRPO} \label{lrpo}

\subsubsection{Term Ordering ({\sc lrpo})}

The {\em lexicographic recursive path ordering} ({\sc lrpo}, or
{\sc rpo} with status) \cite{termination,rta-85,rrl} is a method for
comparing terms.  The important theoretical property of {\sc lrpo} is
that it is a {\em termination ordering}.  That is, let $R$ be a set of
demodulators in which in each demodulator, the left side is
{\sc lrpo}-greater than the right side; then demodulation (applying the
demodulators left to right) is guaranteed to terminate.

To use {\sc lrpo} one typically uses the \verb:lex: command (Sec.
\ref{symbol-commands}) to assign an ordering on constant and function symbols.
If the \verb:lex: command is not present, \otter\ assigns an ordering
(which is frequently ineffective).
(\otter\ uses a total ordering on symbols that is fixed at input time.
Other implementations of {\sc lrpo} use partial orderings or dynamically
changing orderings.)

With respect to {\sc lrpo}, function symbols can have either
{\em left-to-right status} (the default) or {\em multiset status}.
The command {\tt lrpo\_multiset\_status({\it symbol\_list})} gives
symbols multiset status.

{\sc Lrpo} comparison is used when orienting equality literals,
deciding whether an equality should be a demodulator or an
{\sc lrpo}-dependent demodulator, and deciding whether to apply an
{\sc lrpo}-dependent demodulator.  {\sc Lrpo} comparison is never used when
evaluating the functions/predicates that perform lexical comparison
(\verb:$LLT:, \verb:$LGT:, etc.).

\subsubsection{Orienting Equalities ({\sc lrpo})} \label{orient-lrpo}

If the flag \verb:order_eq: is set and if one argument of the equality
literal (positive or negative) is greater in the {\sc lrpo} order, the
greater argument is placed on the left side.  This rule applies to input
demodulators, to inferred clauses, and, if the flag
\verb:process_input: is set, to input \verb:usable: and \verb:sos:
clauses.

\subsubsection{Determining Dynamic Demodulators ({\sc lrpo})} \label{dynamic-lrpo}

If the flag \verb:dynamic_demod: is set, \otter\ attempts to make all
equalities into demodulators (\verb:dynamic_demod_all: is ignored when
\verb:lrpo: is set).  If $\alpha\succ\beta$ in the {\sc lrpo} order, the
derived equality becomes a demodulator  ($\alpha$ is not {\sc
lrpo}-less-than $\beta$, because orienting has already occurred).  If
\verb:dynamic_demod_lex_dep: is set, if neither argument is {\sc
lrpo}-less-than the other, and if every variable that occurs in $\beta$
also occurs in $\alpha$, the derived equality becomes an {\sc
lrpo}-dependent demodulator.

\subsubsection{{\sc lrpo}-dependent Demodulation ({\sc lrpo})} \label{lex-dep-lrpo}

An {\sc lrpo}-dependent demodulator is allowed to rewrite a term if and
only if its application produces an {\sc lrpo}-less-than term.

\subsection{Knuth-Bendix Completion} \label{kb}

The Knuth-Bendix completion procedure \cite{knuth-bendix} attempts to
transform a set $E$ of equalities into a terminating, canonical set of
rewrite rules (demodulators).  If it is successful, the resulting set
of rewrite rules, a {\em complete set of reductions}, is a decision
procedure for equality of terms in the theory $E$.  There are many
variations and refinements of the Knuth-Bendix procedure.

Setting either of the flags
\verb:anl_eq: or
\verb:knuth_bendix:
causes \otter\ to automatically
alter a set of options so that its search will behave like a
Knuth-Bendix completion procedure.  If \otter's search stops because
its \verb:sos: list is empty, and if certain other conditions are met,
then the resulting set of equalities should be a complete set of reductions.
(\otter\ was not designed to implement a completion
procedure, and it has not been optimized for completion.)

\noindent {\em Conjecture}.  If (1) the set $E$ of equalities, along
with \verb:x=x:, is input in list \verb:sos:,
(2) flag \verb:anl_eq: is set, (3) other options that are changed
from the defaults do not affect the search, (4) \otter\ stops with
``sos empty'', and (5) other than \verb:x=x:, the final \verb:usable: list is
the same as the final \verb:demodulators: list, then the \verb:demodulators:
list is a complete set of reductions for $E$.

Here is an input file that causes \otter\ to search for and quickly
find a complete set of reductions for free groups.  Note that the
predeclared (right associative) infix operator \verb:*: is used.

{\small
\begin{verbatim}
    set(anl_eq).
    set(print_lists_at_end).
    lex([e, _*_, g(_)]).

    list(sos).
    x = x.
    e*x = x.          % left identity
    g(x)*x = e.       % left inverse
    (x*y)*z = x*y*z.  % associativity
    end_of_list.
\end{verbatim}
}
\noindent
The critical issue in most applications of the Knuth-Bendix completion
procedure is the choice of ordering scheme and/or the specific
ordering on symbols.  Note, in this case, that if the \verb:lex:
command is absent, the default symbol ordering suffices because
it is essentially the same as the one specified.

The \verb:anl_eq: flag
is also very useful when trying to prove
equational theorems.  When using
\verb:anl_eq: to search for proofs, we are not bound by the
conditions listed in the above claim; in fact, we usually apply
additional strategies such as limiting the size of retained
equalities, being more selective about making equalities into
demodulators, and disabling {\sc lrpo} ordering.

With the following input file, \otter\ uses the \verb:anl_eq:
option to prove the difficult half of a group theory theorem of Levi:
{\it The commutator operation is associative if and only if the
commutator of any two elements lies in the center of the group.}  (A
textbook proof can be found in \cite{kurosh}.)  Note that,
contrary to common practice, the symbol
order does not cause the definition of the commutator operation
\verb:h(_,_): to be used as a rewrite rule to eliminate commutator
expressions in \verb:h:.  Note also that weight templates are used to
eliminate clauses containing terms with particular structures; this
decision is purely heuristic, derived from experimentation and
intuition.  \otter\ finds a proof in about a minute
and uses about 6 megabytes of memory.

{\small
\begin{verbatim}
    set(anl_eq).
    lex([a,b,c,e,h(_,_),f(_,_),g(_)]).
    assign(max_weight, 20). assign(pick_given_ratio, 5).
    clear(print_kept).
    clear(print_new_demod). clear(print_back_demod).

    list(usable).
    x = x.
    f(e,x) = x.                 % group theory
    f(g(x),x) = e.
    f(f(x,y),z) = f(x,f(y,z)).
    end_of_list.

    list(sos).
    f(g(x),f(g(y),f(x,y))) = h(x,y). % definition of commutator
    h(h(x,y),z) = h(x,h(y,z)).       % commutator is associative
    % Denial: there are two elements whose commutator
    % is not in the center.
    f(h(a,b),c) != f(c,h(a,b)).
    end_of_list.

    weight_list(purge_gen).
    weight(h($(0),f($(0),h($(0),$(0)))), 100).
    weight(h(f($(0),h($(0),$(0))),$(0)), 100).
    weight(h($(0),f(h($(0),$(0)),$(0))), 100).
    weight(h(f(h($(0),$(0)),$(0)),$(0)), 100).
    weight(h($(0),h($(0),h($(0),$(0)))), 100).
    weight(h($(0),f($(0),f($(0),$(0)))), 100).
    weight(h(f($(0),f($(0),$(0))),$(0)), 100).
    end_of_list.
\end{verbatim}
}

\section{Evaluable Functions and Predicates ({\tt \$SUM}, {\tt \$LT}, $\ldots$)} \label{eval}

\otter\ can be used in a ``programmed'' mode that is quite different
from normal refutational theorem proving.  When using the programmed
mode, one generally has in mind a particular method for solving a
problem; and when writing clauses for the programmed mode, one
generally knows exactly how they will be used by \otter.

The programmed mode frequently involves a set of evaluable function
and predicate symbols known as the \verb:$:-symbols (because each
starts with \verb:$:).  Examples are \verb:$SUM: and \verb:$LT:
for integer arithmetic and \verb:$AND: for Boolean operations.
%$

The evaluable symbols operate on five types of \otter\ term: integer
constants, floating-point constants, bit-string constants,
the Boolean constants \verb:$T: and
\verb:$F:, and arbitrary terms.  The symbols that evaluate to type
Boolean can occur either as function symbols or as predicate symbols.
The integer, bit, and floating-point operations behave the same as
the underlying C operations applied to the data types ``long int'',
``unsigned long int'', and ``double'', respectively.  Table \ref{eval-tab}
lists the evaluable functions and predicates by type.

\begin{table}[ht] \centering
\caption{Evaluable Functions and Predicates}  \label{eval-tab}
\begin{tabular}{|l|l|}
\hline
$int \times int \rightarrow int$ & \verb:$SUM:, \verb:$PROD:, \verb:$DIFF:, \verb:$DIV:, \verb:$MOD: \\
\hline
$int \times int \rightarrow bool$ & \verb:$EQ:, \verb:$NE:, \verb:$LT:, \verb:$LE:, \verb:$GT:, \verb:$GE: \\
\hline
\hline
$float \times float \rightarrow float$ & \verb:$FSUM:, \verb:$FPROD:, \verb:$FDIFF:, \verb:$FDIV: \\
\hline
$float \times float \rightarrow bool$ & \verb:$FEQ:, \verb:$FNE:, \verb:$FLT:, \verb:$FLE:, \verb:$FGT:, \verb:$FGE: \\
\hline
\hline
$bits \times bits \rightarrow bits$ & \verb:$BIT_AND:, \verb:$BIT_OR:, \verb:$BIT_XOR: \\
\hline
$bits \times int \rightarrow bits$ & \verb:$SHIFT_LEFT:, \verb:$SHIFT_RIGHT:  \\
\hline
$bits \rightarrow bits$ & \verb:$BIT_NOT: \\
\hline
$int \rightarrow bits$ & \verb:$INT_TO_BITS: \\
\hline
$bits \rightarrow int$ & \verb:$BITS_TO_INT: \\
\hline
\hline
$\rightarrow bool$ & \verb:$T:, \verb:$F: \\
\hline
$bool \times bool \rightarrow bool$ & \verb:$AND:, \verb:$OR: \\
\hline
$bool \rightarrow bool$ & \verb:$TRUE:, \verb:$NOT: \\
\hline
$bool \times term \times term \rightarrow term$ & \verb:$IF: \\
\hline
\hline
$term \times term \rightarrow bool$ (lexical) & \verb:$ID:, \verb:$LNE:, \verb:$LLT:, \verb:$LLE:, \verb:$LGT:, \verb:$LGE: \\
\hline
$term \times term \rightarrow bool$ (other) & \verb:$OCCURS:, \verb:$VOCCURS:, \verb:$VFREE:, \verb:$RENAME: \\
\hline
$term \rightarrow bool$ & \verb:$ATOMIC:, \verb:$INT:, \verb:$BITS:, \verb:$VAR:, \verb:$GROUND:  \\
\hline
\hline
$\rightarrow int$ & \verb:$NEXT_CL_NUM:, \verb:$UNIQUE_NUM:  \\
\hline
\end{tabular}
\end{table}

Additional notes on the operations (unless otherwise stated, the
term in question evaluates if all arguments demodulate/evaluate to the
appropriate type):
\begin{itemize}
\item
$int \times int \rightarrow int$.
The symbol \verb:$SUM: is addition, \verb:$PROD: is multiplication,
\verb:$DIFF: is subtraction, \verb:$DIV: is integer division,
and \verb:$MOD: is remainder.
\item
$float \times float \rightarrow float$.
These operations are analogous to the integer operations except
that there is no floating-point remainder operation.
The syntax of floating-point numbers is described in Sec.~\ref{float}
\item
$int \times int \rightarrow bool$.
These are the ordinary relational operations on integers.
The symbol \verb:$EQ: is $=$, \verb:$NE: is $\neq$,
\verb:$LT: is $<$, \verb:$LE: is $\leq$,
\verb:$GT: is $>$,  and \verb:$GE: is $\geq$.
\item
$bits \times int \rightarrow bits$.  The shift operations
\verb:$SHIFT_LEFT: and \verb:$SHIFT_RIGHT: shift the first
argument by the number of places given by the second argument.
\item
$bits \times bits\rightarrow bits$.
The symbols \verb:$BIT_AND:, \verb:$BIT_OR:, and \verb:$BIT_XOR:
are the bitwise conjunction, disjunction, and exclusive-or operations.
\item
$bits \rightarrow bits$.
The symbol \verb:$BIT_NOT: is the one's complement operation on bit strings.
\item
$int \rightarrow bits$.
The symbol \verb:$INTS_TO_BITS: translates a decimal integer
to a bit string.
\item
$bits \rightarrow int$.
The symbol \verb:$BITS_TO_INT: translates a bit string to the corresponding
decimal integer.
\item
$\rightarrow bool$.
The symbols \verb:$T: and \verb:$F: represent {\em true} and {\em false}.
When they appear as literals or atomic formulas in clauses, the
clauses are simplified as appropriate.
% \item
% $bool \times bool \rightarrow bool$.
\item
$bool \rightarrow bool$.
The symbol \verb:$TRUE: is essentially a ``no operation'' on Boolean
constants.  It is used to trick hyperresolution into evaluating literals
(see below).

\item
$bool \times term \times term \rightarrow term$.  The \verb:$IF:
function is the {\it if-then-else} operator.  When inside-out (the
default) demodulation encounters a term {\tt \$IF({\it condition},
$t_1$, $t_2$)}, demodulation takes a path different from its normal
inside-out behavior.  The term {\it condition} is demodulated
(evaluated); if the result is \verb:$T:, the value of the \verb:$IF:
term is the result of demodulating $t_1$; if the result is \verb:$F:,
the value of the \verb:$IF: term is the result of demodulating $t_2$;
if the result is neither \verb:$T: nor \verb:$F:, demodulation returns
to its normal behavior.  Note that if the condition evaluates to a
Boolean value, demodulation deviates from its inside-out behavior,
because just one of $t_1$ and $t_2$ is demodulated.  (If demodulation
were always outside-in, \verb:$IF: would not need to be built in because it
could be efficiently defined with the two demodulators
\verb:if($T,x,y)=x: and \verb:if($F,x,y)=y:.)

\item
$term \times term \rightarrow bool$ (lexical).
These operations are analogous to the six operations in $int \times int
\rightarrow bool$ except that the comparisons are lexical instead of
arithmetic.  The symbol \verb:$ID: tests identity of terms.
The lexical comparison is the same as in lex-dependent
demodulation; in particular, the flag \verb:lex_order_vars: (Secs.
\ref{eq-flags} and \ref{lex-order}) is consulted during these operations.

\item
$term \times term \rightarrow bool$ (other).
The term \texttt{\$OCCURS($t_1$,$t_2$)} is true if $t_1$ is a subterm of
$t_2$, including the case when they are the same.
The term \texttt{\$VOCCURS($t_1$,$t_2$)} is true if $t_1$ is a variable
that occurs in $t_2$.
The term \texttt{\$VFREE($t_1$,$t_2$)} is true if $t_1$ is a variable
that does not occur in $t_2$.
The term \texttt{\$RENAME($t_1$,$t_2$)} is true if $t_1$ and $t_2$
have the same structure; that is, if we rename all variables to
$x$, the terms are identical.

\item
$term \rightarrow bool$.
A term is \verb:$ATOMIC: iff it is a constant (including integer and
bit string),
a term is a \verb:$INT: iff it is an integer,
a term is a \verb:$BITS: iff it is a string of \{0,1\},
a term is a \verb:$VAR: iff it is a (unbound) variable, and
a term is a \verb:$GROUND: iff it does not contain any variables.

\item
$ \rightarrow int$.  The term \verb:$NEXT_CL_NUM: (no arguments)
evaluates to the next integer that will be assigned as a clause
identifier (this is useful for placing the ID of a clause within the
clause).  A sequence of calls to \verb:$UNIQUE_NUM: (no arguments)
returns $[1,2,3,\cdots]$.

\end{itemize}

Evaluation occurs as part of the demodulation process.  In
particular, if demodulation comes across an evaluable term, say
\verb:$SUM(2,3):, it tries to convert the arguments into the
appropriate type (integers for \verb:$SUM:); then if the arguments have the
correct type, it rewrites the term to the result of the operation, in
this case, just as if the demodulator \verb:$SUM(2,3)=5: had been
present.
The evaluation mechanisms, along with ordinary demodulation, form a
reasonably complete (although not particularly speedy or convenient)
equational programming subsystem.

Evaluation/demodulation can also occur, in a very particular way,
during hyperresolution.  (Recall that hyperresolution takes a clause,
the {\em nucleus}, with some negative literals, the conditions, and
resolves each negative literal with a positive clause, producing a
clause with no negative literals.)  Just as evaluation during
demodulation can be thought of as rewriting with an implicit
demodulator, evaluation during hyperresolution can be thought of
resolving with the implicit positive unit clause \verb:$T:
%$
(meaning ``true'').  The mechanism is this: if hyperresolution encounters a
negative literal that has an evaluable predicate symbol, then it
demodulates the atom (the literal without the sign); if the result of
the demodulation is \verb:$T:,
%$
then the literal is considered to have been resolved.

During hyperresolution, demodulation/evaluation is triggered by the
presence of an evaluable literal.  In many cases, however, the user
defines a Boolean function to trigger the mechanism.
Consider the following definition of list membership, written as demodulators:

{\small
\begin{verbatim}
    member(x,[]) = $F.
    member(x,[y|z]) = $IF($ID(x,y),
                           $T,
                           member(x,y)).
\end{verbatim}
}
\noindent
Because the symbol \verb:member: is not evaluable, the
demodulation/evaluation mechanism will not be activated;
however, the unary evaluable predicate \verb:$TRUE:
%$
 can be used in the following way to trigger demodulation/evaluation.
\begin{verse} \small
{\tt $-L_1$ | $\cdots$ | -\$TRUE(member({\it element}, {\it list})) | $\cdots$ | $-L_n$ | $M$}.
\end{verse}
Evaluable functions and predicates are useful to implement
forward-chaining rule-based systems, for example, state-space search
problems (Sec.~\ref{eval-examples}).

Hyperresolution operates on the conditions (negative literals) in
order, left to right.  (The preceding sentence is not quite true
because the first step is typically resolution of a positive given
clause with any one of the conditions,
but for this paragraph, we may assume that it is true.)
If a literal resolves or evaluates, the next literal is considered.
If nothing more can be done with a literal, then hyperresolution
backtracks to the preceding literal in search of an alternative.
When a nucleus
contains evaluable conditions, the order of the conditions is
important both for efficiency and for actually deriving
hyperresolvents.  Evaluable conditions typically have variables
that must be instantiated when nonevaluable literals are resolved.
If an evaluable literal is too far to the left, its variables
will not be sufficiently instantiated when hyperresolution encounters it,
evaluation will fail, and possible paths to hyperresolvents will be
blocked.  
If an evaluable literal is too far to the right, then hyperresolution can
explore many paths that are sure to fail.

\noindent {\em Technical Note and Advice.}  The evaluable symbols
are an add-on feature rather than an integral part of \otter.  In
particular, the objects that are manipulated (integers, bit strings, etc.)
in most cases are stored by \otter\ as character strings rather than
as the appropriate data type.  To evaluate a term, say \verb:$SUM(2,3):,
%$
\otter\ must find the strings \verb:"2": and \verb:"3": in a hash table,
translate them to integers, add them, translate the result to the
string \verb:"5":, then look up \verb:"5":, and possibly insert it into
the hash table.  This procedure is obviously much slower than it needs
to be.  If a problem requires a hundred million
evaluations, the user should consider using something else, including
writing a special-purpose C program.

\noindent {\em Warning 1.}  The evaluable symbols should not be thought of as
theories ``built in'' to \otter.
As theories, they are very incomplete, and \otter\ uses them only in
very constrained ways.

\noindent {\em Warning 2.}  Ordinary resolution inference rules
(e.g., \verb:binary_res:, \verb:hyper_res:, \verb:ur_res:)
never apply to evaluable literals.

\subsection{Using More Natural Expressions for Evaluation} \label{make-eval}

Writing complex evaluable expressions with \verb:$:-symbols can be
quite tedious. Therefore, a feature was added that allows more natural
expressions.  The command \verb:make_evaluable: copies the evaluation
properties from a \verb:$:-symbol to any other symbol of the same
arity.  The form of the command is
\begin{verse} \small
{\tt make\_evaluable(\em any-symbol\tt ,\em evaluable-symbol\tt ).}
\end{verse}
The symbols in the command are given dummy arguments to specify the
arity.  The following list contains typical examples for integer
arithmetic (assuming the symbols on the left are already known to be
infix).
\begin{verse} \small
\verb:make_evaluable(_+_, $SUM(_,_)).:\\
\verb:make_evaluable(_-_, $DIFF(_,_)).:\\
\verb:make_evaluable(_>_, $GT(_,_)).:\\
\verb:make_evaluable(_>=_, $GE(_,_)).:
\end{verse}
\noindent

\noindent
{\em Warning 1.}  If a binary symbol that is recognized by
paramodulation or demodulation as an equality symbol is
given evaluation properties, it will no longer be
recognized by paramodulation or demodulation.  For example,
if the command
\verb:make_evaluable(_=_, $EQ(_,_)):
%$
is issued, paramodulation
and demodulation will not recognize \verb:a=b: as an equality.
The convention is to use \verb:==: for evaluation.

\noindent
{\em Warning 2.}  This is not an ``alias'' mechanism;
the symbols remain distinct for unification, matching,
and identity testing.

\subsection{Evaluation Examples} \label{eval-examples}

\paragraph{Equational Programming.}
The evaluable functions and predicates enable the use of equalities
with demodulation as a general-purpose equational programming
language.  Here are some examples.

{\small
\begin{verbatim}
  gcd(x,y) =    % greatest common divisor for nonnegative integers
     $IF($EQ(x,0),
         y,
         $IF($EQ(y,0),
             x,
             $IF($LT(x,y),
                 gcd(x,$DIFF(y,x)),
                 gcd(y,$DIFF(x,y))))).

  factorial(x) =    % factorial for nonnegative integers
     $IF($EQ(x,0),
         1,
         $PROD(x,factorial($DIFF(x,1)))).

  quick_sort([]) = [].     % naive quicksort
  quick_sort([x|y]) = append(quick_sort(le_list(x,y)),
                              [x|quick_sort(gt_list(x,y))]).
  le_list(z,[]) = [].
  le_list(z,[x|y]) = $IF($LLE(x,z),
                          [x|le_list(z,y)],
                          le_list(z,y)).
  gt_list(z,[]) = [].
  gt_list(z,[x|y]) = $IF($LGT(x,z),
                          [x|gt_list(z,y)],
                          gt_list(z,y)).
\end{verbatim}
}

\paragraph{A State-Space Search.}
Here is a complete \otter\ input file for a simple state-space search.

{\small
\begin{verbatim}
% We have a 3-gallon jug and a 4-gallon jug, both empty,
% and a well. Our goal is to have exactly 2 gallons in the
% 4-gallon jug.  We can fill a jug from the well, empty a
% jug onto the ground, and carefully pour water from one
% jug into the other.
%
% j(m, n) is the state in which the 3-gallon jug contains
% m gallons, and the 4-gallon jug contains n gallons.
 
set(hyper_res).

make_evaluable(_+_, $SUM(_,_)).
make_evaluable(_-_, $DIFF(_,_)).
make_evaluable(_<=_, $LE(_,_)).
make_evaluable(_>_, $GT(_,_)).

list(usable).
-j(x, y) | j(3, y).  % fill the 3-gallon jug
-j(x, y) | j(0, y).  % empty the 3-gallon jug
-j(x, y) | j(x, 4).  % fill the 4-gallon jug
-j(x, y) | j(x, 0).  % empty the 4-gallon jug
-j(x, y) | -(x+y <= 4) | j(0, y+x).     % small -> big; it fits
-j(x, y) | -(x+y >  4) | j(x- (4-y),4). % small -> big, until full
-j(x, y) | -(x+y <= 3) | j(x+y, 0).     % big -> small; it fits
-j(x, y) | -(x+y >  3) | j(3,y- (3-x)). % big -> small, until full

-j(x, 2).  % goal state --- 4-gallon jug containing 2 gallons
end_of_list.

list(sos).
j(0, 0).   % initial state --- both jugs empty
end_of_list.
\end{verbatim}
}

\section{Weighting} \label{weighting}

\otter\ recognizes four lists of weight templates.
(See Sec.~\ref{input-weight} for input of weight template lists.)

\begin{description}
\item[{\tt weight\_list(pick\_given).}]
This list is used for selection of given clauses from list \verb:sos:.
When the weight of a clause is printed, it is the \verb:pick_given: weight.
\item[{\tt weight\_list(purge\_gen).}]
This list is used in conjunction with the \verb:max_weight: parameter
to discard generated clauses.
\item[{\tt weight\_list(pick\_and\_purge).}]
In many cases, one can use the same weighting strategy for
both selecting given clauses and purging generated clauses.
The \verb:pick_and_purge: list serves the purposes of both the
\verb:pick_given: and the \verb:purge_gen: lists.  If the
\verb:pick_and_purge: list is present, then neither the
\verb:pick_given: nor the \verb:purge_gen: list may be present.
\item[{\tt weight\_list(terms).}]
This list is for calculating the weight of terms when using the
weight-lex-order (Sec.~\ref{lex-order}) to compare terms.
This occurs when the flag \verb:lrpo: is clear when orienting
equality literals (Secs. \ref{orient} and \ref{dynamic}).
\end{description}

\subsection{Weighing Clauses and Literals}

The weight of a clause is always the sum of the weights of its literals
(excluding any answer literals).
The weight of a positive literal is the weight of its atom.
The weight of a negative literal is the weight of its atom plus the
value of the \verb:neg_weight: parameter (Sec.~\ref{misc-parms}).

\subsection{Weighing Atoms and Terms}

Atoms and terms are weighed top-down.  To weigh a given term, \otter\
searches the appropriate weight list (in the order input)
for the first matching template.  If a match is found, then the
subterms of the given term that match the integers in the template are
weighed.  The weight of the given term is the sum of the products of
each integer and the weight of its corresponding subterm, plus the
second argument of the weight template.  For example, the template
\begin{verse}
\verb:weight(f(g($(2)),$(-3)), -50).:
\end{verse}
matches the given term
\begin{verse}
\verb:f(g(h(a)),f(b,x)).:
\end{verse}
Let $wt(t)$ be the weight of term or atom $t$. Then
\begin{verse}
$wt(\mbox{\tt f(g(h(a)),f(b,x))}) = 2*wt(\mbox{\tt h(a)}) + (-3)*wt(\mbox{\tt f(b,x)}) + (-50)$.
\end{verse}
If a matching weight
template is not found, then the weight of the given term is 1 plus the sum of
the weights of the subterms.  (See the flags \verb:atom_wt_max_args:
and \verb:term_wt_max_args:, Sec.~\ref{misc-flags}, for overrides.)
Note that this weighting scheme implies that if no weight templates
are present, the default weight of a term or atom is the number of
variable, constant, function, and predicate symbols (the symbol count).

Variables in weight templates are generic.
A variable in a weight template will match any variable, and only a variable,
in the given term.
As a consequence, it is never necessary to use different variable names
in a weight template.
For example, \verb:weight(f(x,x),-7): matches the term \verb:f(u,v):,
and \verb:weight(x,32): matches all variables.

{\em Warning.} The two occurrences of symbol \verb:f: in the term
\verb:f(f,x): are treated by \otter\ as different symbols because they
have different arities.  The weight template \verb:weight(f, 0):
applies to the second occurrence but not to the first.

The default weight of an answer literal is 0, but templates can
be used to assign weights to answer literals.  The parameter
\verb:neg_weight: never applies to answer literals.

If one wishes to have a weight template containing a Skolem function or
constant that is generated by \otter, one must first make a short
trial run to find out how the formulas are Skolemized, then return
to the input file and insert the weight list containing the
Skolem symbol {\it after} the formula lists.

\subsection{Containment Weight Templates}

Term weighting has an additional feature that allows
the user to specify terms that \emph{contain} particular terms.
This is done with a unary function symbol \verb:$dots(:$t$\verb:):.
If \verb:$dots(:$t$\verb:): occurs in a weight template,
it will match any term that contains a term that
matches t.  This is very useful for discarding ``bad'' clauses.
Here is part of an output file that illustrates this feature.

{\small
\begin{verbatim}
  list(sos).
  1 [] p(f(g(g(g(g(g(h(h(h(h(j(b)))))))))))).
  2 [] p(F(g(g(h(g(H(g(h(g(g(g(B)))))))))))).
  3 [] p(f3(g(h(a)),g(g(b)),h(h(c)))).
  end_of_list.
  
  weight_list(pick_given).
  weight(f($dots(j($(5)))),100).
  weight(F($dots(H($dots(B)))),1000).
  weight(f3($dots(a),$dots(b),$dots(c)),2000).
  end_of_list.
  
  ======= end of input processing =======
  =========== start of search ===========
  
  given #1: (wt=106) 1 [] p(f(g(g(g(g(g(h(h(h(h(j(b)))))))))))).
  given #2: (wt=1001) 2 [] p(F(g(g(h(g(H(g(h(g(g(g(B)))))))))))).
  given #3: (wt=2001) 3 [] p(f3(g(h(a)),g(g(b)),h(h(c)))).
\end{verbatim}
}
%$

\section{Answer Literals} \label{answer}

The main use of answer literals is to record, during a search for a
refutation, instantiations of variables in input clauses.  For
example, if the theorem under consideration states that an object
exists, then the denial of the theorem contains a variable, and an
answer literal containing the variable can be appended to the denial.
If a refutation is found, then the empty clause has an answer literal
that contains the object whose existence has just been proved.

Any literal whose predicate symbol starts with \verb:$ans:,
\verb:$Ans:, or \verb:$ANS: is an answer literal.  Most
%$
routines---including the ones that count literals and decide whether a
clause is positive or negative---ignore any answer literals.  The
inference rules insert, into the children, the appropriate instances
of any answer literals in the parents.  If factoring is enabled,
\otter\ {\it does} attempt to factor answer literals.

\section{The Passive List} \label{passive}

Either clauses or formulas can be input to list \verb:passive:.
After input, the passive
list is fixed for the rest of the run.  Clauses in the passive
list are used for exactly two purposes: forward subsumption and
unit conflict.  If forward subsumption is enabled, a newly
generated clause will be deleted if it is subsumed by any clause in
\verb:usable:, \verb:sos:, or \verb:passive:, and newly kept
unit clauses are checked for unit conflict against unit clauses in
\verb:usable:, \verb:sos:, or \verb:passive:.

The passive list has been most useful for monitoring the progress of a
search.  Suppose we are trying to prove a difficult theorem, we have
some lemmas in mind, and we would like to know whether \otter\ has proved
the lemmas.  Then denials of the lemmas can be placed in the passive
list, and \otter\ will report proofs if it proves any lemmas, but the
denials of the lemmas will not interfere with the search for the main
theorem.  (Recall that an appropriate value must be assigned to
\verb:max_proofs:; otherwise \otter\ will stop at the first proof.)

\section{Clause Attributes} \label{attributes}

Attributes can be attached to clauses.  This feature was introduced
at the same time as the hints strategy (Sec.~\ref{hints}),
and all of the current attributes are specifically for the hints strategy.
In case some future enhancements of \otter\ will use attributes,
the general attribute mechanism is given here.

Each attribute is identified by a name, and each attribute
has a type.  (Users cannot introduce new attributes---they
are built into the code of \otter.)
The attribute types are integer, string, and term.
Attributes are attached to clauses with
the operator ``\verb:#:'', and must appear after all literals.

For example, if
attribute \verb:a1: has type integer,
attribute \verb:a2: has type string, and
attribute \verb:a3: has type term, then
a user can write a clause with attributes as follows.

{\small
\begin{verbatim}
  f(x,y)!=f(x,z) | y=z # a1(23) # a2("left cancel") # a3(g(b)).
\end{verbatim}
}

\section{The Hints Strategy} \label{hints}

The \emph{hints strategy} can be used if the user has a set of
clauses that might be relevant to finding a proof.
The clauses, called \emph{hints}, do not necessarily hold
in the theory being explored, and they are not used for making inferences.
Hints are used only as a heuristic for guiding the search, in particular,
in selecting the given clauses and in deciding whether to keep
derived clauses.

The main function of the hints strategy is to adjust the pick-given
weight of clauses.  The user can specify, for example, that any derived
clause that matches a hint should have its pick-given weight reduced
by 1000.  In addition, the user can specify, with an attribute on a hint,
how that hint should be used to adjust the pick-given weight of
clauses that match the hint.  Clauses can match hints in several
ways as specified by (ordinary) parameters and by attributes on hints.

Because of the distinction between the pick-given and purge-gen
weights of clauses, and because the hints mechanism affects only
the pick-given weight, several additional flags exist.  If a clause
matching a hint is derived, one typically wants it to be kept so
that it can be selected as a given clause.  However, the clause
may be discarded by the \verb:max_weight: parameter.
To address this problem, the flags
\verb:keep_hint_subsumers: and \verb:keep_hint_equivalents: say
that the \verb:max_weight: parameter should be ignored for
all clauses that match hints in those ways (details are
in the following subsections).

The hints strategy was introduced by Bob Veroff, who implemented
it in a previous version of \otter\ and used it in many applications
\cite{veroff:hints,veroff:sketches}.
\otter\ currently has two separate
hints mechanisms, named ``hints'' and ``hints2'',
both derived from Veroff's methods and ideas.  The first is more
general, and the second is much faster.

\subsection{Hints (the General Version)}

The hint clauses are given in one or more lists.  All of the
lists must be named ``hints'' as in the following example.

{\small
\begin{verbatim}
    list(hints).
    -p(x) | q(x) | r(x).
    end_of_list.
\end{verbatim}
}
\noindent
A clause $C$ can match a hint $H$ in three ways.
\begin{enumerate}
\item $C$ subsumes $H$.
This is referred to as ``bsub'', in analogy to back subsumption.
\item $C$ is subsumed by $H$.
This is referred to as ``fsub'', in analogy to forward subsumption.
\item $C$ is equivalent to $H$.
\end{enumerate}
Six parameters and two flags determine the behavior of the hints mechanism.

\noindent
\verb:assign(equiv_hint_wt,:$n$\verb:):.  Default \maxint, range [$-$\maxint ..\maxint ].
If $n \neq$ \maxint,
clauses that are equivalent to a hint receive a pick-given weight of $n$.

\noindent
\verb:assign(equiv_hint_add_wt,:$n$\verb:):.  Default 0, range [$-$\maxint ..\maxint ].
Clauses that are equivalent to a hint have $n$ added to their ordinary
pick-given weight.

\noindent
\verb:assign(fsub_hint_wt,:$n$\verb:):.  Default \maxint, range [$-$\maxint ..\maxint ].
If $n \neq$ \maxint,
clauses that are subsumed by a hint receive a pick-given weight of $n$.

\noindent
\verb:assign(fsub_hint_add_wt,:$n$\verb:):.  Default 0, range [$-$\maxint ..\maxint ].
Clauses that are subsumed by a hint have $n$ added to their ordinary
pick-given weight.

\noindent
\verb:assign(bsub_hint_wt,:$n$\verb:):.  Default \maxint, range [$-$\maxint ..\maxint ].
If $n \neq$ \maxint,
clauses that subsume a hint receive a pick-given weight of $n$.

\noindent
\verb:assign(bsub_hint_add_wt,:$n$\verb:):.  Default 0, range [$-$\maxint ..\maxint ].
Clauses that subsume a hint have $n$ added to their ordinary
pick-given weight.

A clause can match more than one hint, and a clause can match a hint
in more than one way, so the order of operations is relevant.
The rules are as follows.  (1) The first hint (as given in the input
file) that matches the clause is used.  (2) Equivalence is tested
first, then fsub, then bsub.  (3) Within match type (e.g., bsub), both
types of weight adjustment can be applied (e.g.,
\verb:bsub_hint_wt: and \verb:bsub_hint_ad_wt:).

The hint-adjustment parameters can be overridden by attributes
on individual hints.  (This feature can be used, for example, if some hints
are more important than others.)  The attribute names for hints
correspond to the six parameters listed above.  For example, if
the parameter \verb:bsub_hint_add_wt: is set to -1000, that
value can be overridden for a particular hint by giving it
an attribute as follows.

{\small
\begin{verbatim}
    f(x,f(x,f(x,y)))=f(x,y) # bsub_hint_add_wt(-2000).
\end{verbatim}
}

The following two flags were introduced because the
hints mechanism adjusts the pick-given weight of clauses
and does not affect the purge-gen weight of clauses.

\noindent
Flag \verb:keep_hint_subsumers:.  Default clear.
If this flag is set, then the \verb:max_weight: parameter is
ignored for derived clauses that subsume any hints.

\noindent
Flag \verb:keep_hint_equivalents:.  Default clear.
If this flag is set, then the \verb:max_weight: parameter is
ignored for derived clauses that are equivalent to any hints.

\paragraph{Practical Advice.}  If one has a set of clauses that
one wishes to use as hints, say from a proof of a related theorem,
a good first attempt is to use the following settings.

\begin{verbatim}
    assign(bsub_hint_add_wt, -1000).
    set(keep_hint_subsumers).
\end{verbatim}
If one has more than a few hints, and one wishes to use only the preceding
settings, then we recommend using hints2 (the fast version).

\subsection{Hints2 (the Fast Version)}

The general hints mechanism described in the preceding paragraphs
can be very slow if there are many hints, because each hint
clause is tested until a match is found.  The fast hints mechanism
uses indexing to find hints.  To activate the fast hints
mechanism, hint clauses are placed in one or more lists named
``hints2'' as in the following example.

{\small
\begin{verbatim}
    list(hints2).
    -p(x) | q(x) | r(x).
    end_of_list.
\end{verbatim}
}
\noindent
Only one type of matching can be used with hints2---a clause matches
a hint if and only if it subsumes the hint.  Two parameters
and two flags apply to hints2.

\noindent
\verb:assign(bsub_hint_wt,:$n$\verb:):.  Default \maxint, range [$-$\maxint ..\maxint ].
If $n \neq$ \maxint,
clauses that subsume a hint receive a pick-given weight of $n$.

\noindent
\verb:assign(bsub_hint_add_wt,:$n$\verb:):.  Default 0, range [$-$\maxint ..\maxint ].
Clauses that subsume a hint have $n$ added to their ordinary
pick-given weight.

\noindent
Flag \verb:keep_hint_subsumers:.  Default clear.
If this flag is set, then the \verb:max_weight: parameter is
ignored for derived clauses that subsume any hints.

\noindent
Flag \verb:degrade_hints2:.  Default clear.
If this flag is set, Bob Veroff's hint-degradation strategy is applied.
When a hint is first read, its \verb:bsub_hint_add_wt: is associated
with it; this value can come from the parameter or from an attribute.
The hint-degradation strategy says that each time a hint is used
to adjust the weight of a derived clause, its \verb:bsub_hint_add_wt:
is cut in half.  Assuming that it initially has a large negative value,
this strategy makes a hint progressively less important as it matches
more derived clauses.  This strategy was introduced because
(contrary to intuition) many different generalizations of a hint
can be derived.

\subsection{Label Attributes on Hints}
If a hint clause has a label attribute, for example,
\begin{verbatim}
  f(x,f(x,f(x,y)))=f(x,y) # label("hint 32 from proof 12").
\end{verbatim}
and if such a hint is used to adjust the pick-given weight
of a clause $C$, the label is inherited by $C$.  This
feature, which is useful for tracking the application of hints,
applies to both the general and the fast hints mechanisms.

\subsection{Generating Hints from Proofs}

The main source for hints is proofs of related theorems.
(If the goal is to shorten a proof, hints often come from
proofs of the same theorem.)

\noindent
Flag \verb:print_proof_as_hints:.  Default clear.
If this flag is set, then whenever a proof is found,
it is printed as a hints list in a form that can
be input to a subsequent \otter\ job.  The proof
that is printed contains more detail than the ordinary
proofs printed by \otter.  In particular, when the proof
contains demodulation, clauses are printed for each rewrite
step of demodulation.
This flag is independent of the hints mechanism.

\section{Interaction during the Search} \label{interact}

\otter\ has a primitive interactive feature that allows the user to
interrupt the search, modify the options, and then continue the
search.  The interrupt is triggered in two ways: (1) with \otter\ running in
the foreground, the user types the ``interrupt'' character (often {\sc
delete} or control-C), or (2) if the parameter \verb:interrupt_given:
is set to $n$, the search is interrupted after every $n$ given clauses.
When interrupted, \otter\ immediately goes into a simple loop to read
and execute commands.
The accepted commands are listed in Table \ref{interact-tab}.
\begin{table}[htbf] \centering \small
\caption{Interaction Commands} \label{interact-tab}
\begin{tabular}{ll} \hline
{\tt help.}  &  Give simple help. \\
{\tt set({\it flag-name}).} & Set a flag. \\
{\tt clear({\it flag-name}).}  & Clear a flag.\\
{\tt assign({\it param-name},{\it value}).} & Assign a value to a parameter. \\
{\tt stats.} & Send statistics to std. output and the terminal. \\
{\tt usable.} & Print list \verb:usable: on the terminal. \\
{\tt sos.} & Print list \verb:sos: on the terminal. \\
{\tt demodulators.} & Print list \verb:demodulators: on the terminal. \\
{\tt passive.} & Print list \verb:passive: on the terminal. \\
{\tt fork.} & Fork and run the child process; \\
            & \hspace{1cm} resume parent when child finishes. \\
{\tt continue.} & Continue the search. \\
{\tt kill.} & Send statistics to standard output, and exit. \\
\hline
\end{tabular}
\end{table}

\noindent The following notes elaborate on the interactive feature.
\begin{itemize}
\item 
The flag \verb:interactive_given: (Sec.~\ref{loop-flags})
can be useful with the interactive feature.  For example, if one
thinks the search is going to fail, one can interrupt it, print
list \verb:sos:, set the \verb:interactive_given: flag, then continue,
selecting given clauses interactively.
\item
The \verb:fork: command creates a separate copy, called a {\em child},
of the entire \otter\ process.  Immediately after the fork, the child
is running (waiting for more commands), and the original process, the
{\em parent}, is waiting for the child to finish.  When the child
finishes, the parent resumes (waiting for more commands).  Changes
that the child makes to the clause space, options, and so forth, are not
reflected in the parent; when the parent resumes, it is in exactly the
same state as when the fork occurred.  (The timing statistics are not
handled correctly in child processes; CPU times are from the start of
the current process; wall-clock time is correct; other timings are
not reliable.)
\item
The interactive routine is an area where a user who is also a C
programmer can easily add features.  For example, most of the ordinary
input commands could be made available in the interactive mode.
\item
This kind of interaction can be disabled by using the command
\verb:clear(sigint_interact):.
\end{itemize}
{\em Warning.}  Do not interactively change
any option that affects term or literal indexing.

\section{Output and Exit Codes} \label{output}

\otter\ sends most of its output to ``standard output'', which
is usually redirected by the user to a file; we just call it 
the output file.  The first part of the output file is an echo of
most of the input and some additional information, including
identification numbers for clauses and description of some input
processing.  Comments are not echoed to the output.
The second part of the output file reflects the search.
Various print flags determine what is output.  Given clauses,
generated clauses, kept clauses, and several messages about the
processing of generated and kept clauses can be printed.
Both statistics from the parameter \verb:report: and proofs can
also be printed during the search.  The final part of the output
file lists counts of various events (such as clauses given and
clauses kept) and times for various operations.

Whenever a clause is printed, it is printed with its integer
identifier (ID) and a justification list, which is enclosed in brackets.
Examples:

{\small
\begin{verbatim}
  4  [] -j(x,y)|j(x,0).
  13 [hyper,11,8,eval,demod] j(3,1).
  41 [31,demod] p([a,b,b,c,c,c,d,e,f]).
  14 [new_demod,13] f(y,f(y,f(y,x)))=x.
  71 [back_demod,58,demod,70,14,55,11,34,11] e!=e.
  12 [demod,9] f(a,f(b,f(g(a),g(b))))!=e.
  77 [binary,57.3,30.2] sm|mm| -sl.
  33,32 [para_from,26.1.1,15.1.1.2,demod,21] g(x)=f(x,x).
  36 [hyper,31,2,26,30,unit_del,19,18,20,19] p(k,g(k)).
  4  [factor_simp,factor_simp]
     p(x)|p($f1(x))| -q($f2(y))| -q(y)|p($c6).
  199 [binary,198.1,191.1,factor_simp] q($c14).
\end{verbatim}
}
\noindent
If the justification list is empty, the clause was input.
Otherwise, the {\em first} item in the justification list is 
one of the following.
\begin{description}
\item[An inference rule.]
The clause was generated by an inference rule.  The IDs of the parents
are listed after the inference rule with the given clause ID listed
first (unless \verb:order_history is set):.
\item[A clause identifier.]
The clause was generated by the \verb:demod_inf: rule.
\item[{\tt new\_demod}.]
The clause is a dynamically generated demodulator; it is a copy of the
clause whose ID is listed after \verb:new_demod:.
\item[{\tt back\_demod}.]
The clause was generated by back demodulating the clause
whose ID is listed after \verb:back_demod:.
\item[{\tt demod}.]
The clause was generated by back demodulating an input clause.
\item[{\tt factor\_simp}.]
The clause was generated by factor-simplifying an input clause.
For example, \verb:p(x)|p(a): factor-simplifies to \verb:p(a):.
\end{description}
The sublist $[\mbox{\tt demod}, id_1, id_2, \ldots]$ indicates
demodulation with $id_1, id_2, \ldots$.
The sublist $[\mbox{\tt unit\_del}, id_1, id_2, \ldots]$ indicates
unit deletion with $id_1, id_2, \ldots$.
The symbols \verb:eval: indicates that a literal was ``resolved'' by
evaluation (Sec.~\ref{eval}) during hyperresolution.
The sublist $[\mbox{\tt factor\_simp}, \mbox{\tt factor\_simp}, \ldots]$
indicates a sequence of factor-simplification steps (Sec.~\ref{gen-flags}).

In proofs, some clauses are printed with two (consecutive) IDs.
In such a case, the clause is a dynamically generated demodulator,
and the two IDs refer to different copies of the same clause:
the first ID refers to its use for inference rules, and the second
to its use as a demodulator.

If the flag \verb:detailed_history: is set, then for the inference
rules \verb:binary_res:, \verb:para_from:, and \verb:para_into:,
the positions of the unified literals or terms are listed along
with the parent IDs.  For example, \verb:[binary,57.3,30.2]:
means that the third literal of clause 57 was resolved with
the second literal of clause 30.  For paramodulation, the
``from'' parent is listed as $ID.i.j$, where $i$ is the
literal number of the equality literal, and $j$ (either 1 or 2)
is the number of the unified equality argument; the ``into''
parent is listed as $ID.i.j_1.\cdots.j_n$, where $i$ is the
literal number of the ``into'' term, and $j_1.\cdots.j_n$
is the {\em position vector} of the ``into'' term; for example,
\verb:400.3.1.2: refers to the second argument of the
first argument of third literal of clause 400.  If the flag
\verb:para_all: is set, then the paramodulation positions 
are not listed.

When the flag \verb:sos_queue: is set, the search is breadth first
(level saturation), and \otter\ sends a message to the output file
when given clauses start on a new level.
(Input clauses have level 0, and generated
clauses have level one greater than the maximum of the levels of the 
parents.
Since clauses are given in the order in which they are retained,
the level of given clauses never decreases.)

\paragraph{Exit Codes.}
When \otter\ stops running, it sends an exit code to the operating
system, giving the reason for termination.  The codes are useful when
another program or system calls \otter.  Table \ref{exit-tab}
lists the exit codes.  Note that we do not follow the \unix\
convention of returning zero for normal and nonzero for abnormal
termination.
\begin{table}[htbp] \centering
\caption{Exit Codes}  \label{exit-tab}
\begin{tabular}{cl}
\hline
101 & Input error(s) \\
102 & Abnormal end (compile-time limit or \otter\ bug) \\
103 & Proof(s) found (stopped by \verb:max_proofs):\\
104 & \verb:sos: list empty \\
105 & \verb:max_given: parameter exceeded \\
106 & \verb:max_seconds: parameter exceeded \\
107 & \verb:max_gen: parameter exceeded \\
108 & \verb:max_kept: parameter exceeded \\
109 & \verb:max_mem: parameter exceeded \\
110 & Operating system out of memory \\
111 & Interactive exit \\
112 & Memory error (probable \otter\ bug) \\
113 & A USR1 signal was received \\
114 & The splitting rule terminated with a possible model \\
115 & \verb:max_levels: parameter exceeded \\
\hline
\end{tabular}
\end{table}

\section{Controlling Memory} \label{mem-control}

In many \otter\ searches, the \verb:sos: list accumulates many
clauses that never enter the search, possibly wasting a lot
of memory.  The normal way to conserve memory is to put a maximum
on the weight of kept clauses.  It can be difficult, however, to
find an appropriate maximum.  \otter\ has a feature, enabled
by the command \verb:set(control_memory):, that attempts to 
automatically adjust the maximum.

The memory-control feature operates as follows.  When one third of
available memory (\verb:max_mem: parameter) has been filled, \otter\
assigns or reassigns a maximum weight.  The new maximum, say $n$,
is such that 5\% of all clauses in \verb:sos: have weight $\leq n$.
From then on, at every tenth iteration of the main loop,
\otter\ calculates a prospective new maximum $n'$ in the same way.  If
$n' < n$, then the maximum is reset to $n'$.  The values
$1/3$ and 5\% were determined by trial and error.  Perhaps these values should be 
parameters.

\paragraph{Reducing {\tt max\_weight} on the Fly.}
In many searches, the number of kept clauses grows much faster than
the number of given clauses.  In other words, the list \verb:sos: is
very large, and most of those clauses never participate in the search.
To save memory, one can use the \verb:max_weight: parameter to discard
many of the clauses that will (probably) never become given clauses.

A few searches and proofs show a phenomenon we call the {\it
complexity hump}.  To get a search started, one must use complex
clauses; then one can continue the search using simpler clauses.  That
is, the first few steps in the proof are complex, and the remaining
steps are simpler.  If one needs to carefully conserve memory when a
complexity hump is present, one can use the parameters
\verb:change_limit_after: and \verb:new_max_weight: to change the
value of \verb:max_weight: after a specified number of given clauses.

\noindent
\verb:assign(change_limit_after,:$n$\verb:):.  Default 0, range [0..\maxint ].
If $n$ (the value) is not 0, this parameter has effect.
After $n$ given clauses have been used, the parameter \verb:max_weight:
is automatically reset to the value of the parameter \verb:new_max_weight:.

\noindent
\verb:assign(new_max_weight,:$n$\verb:):.  Default \maxint , range [$-$\maxint ..\maxint ].
See the description of the preceding parameter.

Note that the memory-control feature (Sec.~\ref{mem-control}) can also
address the complexity hump phenomenon.

\section{Autonomous Mode} \label{auto}

If the flag \verb:auto: is set, \otter\ will scan the input
clauses for some simple syntactic properties and decide on inference
rules and a search strategy.  We think of the autonomous mode as
providing a built-in metastrategy for selecting search strategies.
The search strategy that \otter\ selects for a particular set of
clauses is usually refutation complete (except for the flag
\verb:control_memory:), but the user should not expect it to be
especially effective.  It will find proofs for many easy theorems, and
even for cases in which it fails to find a proof, it provides a
reasonable starting point.

In the input file, the command \verb:set(auto): must occur before
any input clauses, and all input clauses must be in list
\verb:usable:; it is an error to place input clauses on any of the
other lists when in autonomous mode.  \otter\ will move some of the
input clauses to \verb:sos: before starting the search.  When \otter\
processes the \verb:set(auto): command, it alters some options,
even before examining the input clauses.  If the user wishes to
augment the autonomous mode by including some ordinary \otter\ commands
(including overriding \otter's choices), the commands should be placed after
\verb:set(auto): and before \verb:list(usable):.

After \verb:list(usable): has been read, \otter\ examines the input
clauses for several syntactic properties and decides which inference
rules and strategies should be used, and which clauses should be moved
to \verb:sos:.  The user cannot override the decisions that \otter\
makes at this stage.

\otter\ looks for the following syntactic properties of the set of
input clauses:
(1) whether it is propositional,
(2) whether it is Horn,
(3) whether equality is present,
(4) whether equality axioms are present, and
(5) the maximum number of literals in a clause.
The program then considers six basic combinations of the properties:
(1) propositional,
(2) equality in which all clauses are units, and
(3--6) the four combinations of \{equality, Horn\}.
To see precisely what \otter\ does for these cases, the 
reader can set up and run some simple experiments.

Please be aware that the autonomous mode reflects
individual experience with \otter; other users would certainly formulate
different metastrategies.  For example, one might prefer UR-resolution
to hyperresolution or in addition to hyperresolution in rich Horn or
nearly-Horn theories, and one might prefer to add few or no dynamic
demodulators for equality theories.

\section{Fringe Features} \label{fringe}

This section describes features that are new, not well tested,
and not well documented. \otter\ is not as robust when using these
features, especially when more than fringe features is being used.

\subsection{Ancestor Subsumption}

\otter\ does not necessarily prefer short or simple proofs---it
simply reports the proofs that it finds.  An option
\verb:ancestor_subsume: extends the concept of subsumption to
include the derivation history, so that if two clause occurrences
are logically identical, the one with fewer ancestors is preferred.
The motivation is to find short proofs.

\noindent
Flag \verb:ancestor_subsume:.  Default clear.  If this flag is set,
the notion of subsumption (forward and back) is replaced with
{\it ancestor-subsumption}.  Clause $C$ ancestor-subsumes clause
$D$ iff $C$ properly subsumes $D$ or if $C$ and $D$ are variants and
$size(ancestorset(C)) \leq size(ancestorset(D))$.

When setting \verb:ancestor_subsume:, we strongly recommend
not clearing the flag \verb:back_subsume:, because doing so can cause
many occurrences of the same clause to be retained and used as given
clauses.

\subsection{The Hot List} \label{hot}

The hot list is a strategy that can be used to emphasize particular
clauses.  It was invented by Larry Wos in the context of
paramodulation, and it has been extended to most of \otter's
inference rules.  To use the strategy, the user simply inputs
one or more clauses in the special list named \verb:hot:.
Whenever a clause is generated and kept by \otter's ordinary
mechanisms, it is immediately considered for inference with
clauses in the hot list.

\paragraph{Which Clauses Should Be Hot?}
Clauses input in the hot list are usually copies of clauses that occur
also in \verb:sos: or \verb:usable:.  They are typically clauses
that the user believes will play a key role in the search
for a proof, for example, special hypotheses.

\paragraph{Managing Hot-List Clauses.}
Input to the hot list
is the same as input to other lists and can be in either clause
or formula form, for example,

{\small
\begin{verbatim}
    list(hot).
    f(x,x) = x.  m(m(x)) = x.
    end_of_list.
\end{verbatim}
}
\noindent
The flag \verb:process_input: has no effect on hot-list clauses;
they are never altered during input.  Hot-list clauses are never
deleted, for example by back subsumption or back demodulation.
Even if a hot-list clause is identical to a clause in another
list, it has a unique identifying number, and proofs that use
hot-list clauses generally refer to two copies (with different
ID numbers) of those clauses.

\paragraph{Hot Inference Rules.}
The inference rules that are
applied to newly kept clauses and hot-list clauses are
the same as the rules in effect for ordinary inference, with
the exceptions \verb:demod_inf:, \verb:geometric_rule:, and
\verb:linked_ur_res:, which are never applied to hot-list clauses.

\paragraph{Applying Hot Inference.}
When hot inference is applied,
the newly kept clause is treated as the given clause, and
the hot list is treated as the usable list.  (Note that the
newly kept clause is not in the hot list, so it will not be
considered for inference with itself, as happens with the
given clause in ordinary inference.)  For inference rules such as
hyperresolution or UR-resolution that can use more than two
parents, {\em all} of the other parents must be in the hot list;
this generally means that the nucleus and other satellites
must be in the hot list.  Hot inference is not applied to 
clauses that are ``kept'' during processing of the input.

\paragraph{Level of Hot Inference (Parameter {\tt heat}).}
To prevent long sequences of hot inferences (i.e., hot
inference applied to a clause generated by hot inference, and so on)
we consider the {\em heat level} of hot inference.  The heat level of
an ordinary inference is 0, and the heat level of a hotly inferred
clause is one more than the heat level of the new-clause parent.  The
parameter \verb:heat:, default 1, range [0..100], is the maximum heat
level that will be generated.  When a clause is printed, its heat
level, if greater than 0, is also printed.

\paragraph{Dynamic Hot Clauses (Parameter {\tt dynamic\_heat\_weight}).}
Clauses can be added to the hot list during a search.  If the
\verb:pick_given: weight of a kept clause is less than or equal to the
parameter \verb:dynamic_heat_weight:, default $-$\maxint , range
[$-$\maxint ..\maxint], then
the clause will be added to the hot list and used for subsequent hot
inference.  Input clauses that are ``kept'' during processing of the
input are never made into dynamic hot clauses.  Dynamic hot clauses
can be added to an empty hot list (i.e., no input hot list).

\subsection{Sequent Notation for Clauses} \label{sequent}

Two flags enable the use of sequent notation for clauses.

\noindent
Flag \verb:input_sequent:.  Default clear.  If this flag is set,
clauses in the input file must be in sequent notation.

\noindent
Flag \verb:output_sequent:.  Default clear.  If this flag is set,
then sequent notation is used when clauses are output.

\noindent Syntax:

\begin{itemize}
\item
All sequent clauses have an arrow.
\item
The negative literals (if any) are written on the left side of the arrow,
are written without the negation sign, and are separated by commas.
\item
The positive literals (if any) are written on the right side of the arrow
and are separated by commas.
\end{itemize}

\noindent Table \ref{sequent-tab} lists some examples.
\begin{table}[ht] \centering
\caption{Examples of Sequent Clauses}   \label{sequent-tab}
\begin{tabular}{l|l}
\hline
Ordinary Clause & Sequent Clause \\
\hline
\verb:-p | -q | -r | s | t: & \verb:p,q,r->s,t: \\
\verb:p(a,b,c): & \verb:-> p(a,b,c): \\
\verb:a!=b: & \verb:a=b ->: \\
\verb:$F: (the empty clause) & \verb:->: \\
\hline
\end{tabular}
\end{table}
%$

Note that \verb:p,q->r,s: is ordinarily thought of as
(\verb:p: {\em and} \verb:q:) {\em implies} 
(\verb:r: {\em or} \verb:s:).

Sequent clauses are treated as (parsed as) a special case because
they can't be made to fit within \otter's ordinary syntax.

\subsection{Conditional Demodulation} \label{cond-demod}

A conditional demodulator has the form
\begin{verse}
{\it condition \tt -> \it equality-literal}.
\end{verse}
The equality is applied as a demodulator if and only if the
instantiated {\it condition} evaluates to \verb:$T:.
%$
The equality of
a conditional demodulator is not subjected on input to being flipped or
to being flagged as a lex-dependent demodulator, and conditional
demodulators are never back demodulated.  In other ways, conditional
demodulators behave as ordinary demodulators.  Examples are
(\verb:member: and \verb:gcd: are defined in Sec.~\ref{eval}.)

{\small
\begin{verbatim}
    $ATOMIC(x) -> conjunctive_normal_form(x)=x.
    member(gcd(4,x),y) -> Equal(f(x,y), g(y)).
    $GT($NEXT_CL_NUM,1000) -> e(x,x) = junk.
\end{verbatim}
}
%$

\subsection{Debugging Searches and Demodulation}

The flag \verb:very_verbose: causes too much output to be used
with large searches.  The following parameters can turn on verbose
output for a segment of the search.

\noindent
\verb:assign(debug_first,:$n$\verb:):.  Default 0, range [0 ..\maxint ].
This parameter is consulted if the flag \verb:very_verbose:
is set.  Verbose output will begin when
a clause is kept and given an identifier of this value.

\noindent
\verb:assign(debug_first,:$n$\verb:):.  Default -1, range [-1 ..\maxint ].
This parameter is consulted if the flag \verb:very_verbose:
is set.  Verbose output will end when
a clause is kept and given an identifier of this value.

\noindent
\verb:assign(verbose_demod_skip,:$n$\verb:):.  Default 0, range [0 ..\maxint ].
This parameter is consulted during demodulation if the flag
\verb:very_verbose: is set.
Verbose output will not occur during the first $n$ rewrites.

\subsection{Special Unary Function Demodulation}

A feature, activated by the \verb:special_unary: command,
allows \otter\ to avoid one of the problems caused by
the lack of associative-commutative matching during demodulation.
The feature is useful when an associative-commutative function and
an inverse are present, as in rings.
Without this feature, the following \verb:lex: command and
demodulators

{\small
\begin{verbatim}
    lex([0,a,b,c,d,e,g(_),f(_,_)]).

    list(demodulators).
    f(x,y) = f(y,x).
    f(x,f(y,z)) = f(y,f(x,z)).
    f(x,g(x)) = 0.
    f(x,f(g(x),y)) = f(0,y).
    f(0,x) = x.
    end_of_list.
\end{verbatim}
}
\noindent
will cause the expression

{\small
\begin{verbatim}
    f(f(f(g(b),a),c),f(b,g(c)))
\end{verbatim}
}
\noindent
to be sorted into

{\small
\begin{verbatim}
    f(a,f(b,f(c,f(g(b),g(c))))).
\end{verbatim}
}
\noindent
One would like \verb:b: and \verb:g(b): to be next to each other so that
they could be canceled by one of the inverse demodulators.
The special-unary feature accomplishes just that.  The command

{\small
\begin{verbatim}
    special_unary([g(x)])
\end{verbatim}
}
\noindent
causes \verb:g: to be ignored during term comparisons, and the
expression will be demodulated to \verb:a:.
The \verb:special_unary: command has no effect if the flag
\verb:lrpo: is set.
{\it This is a highly experimental feature.  Its behavior has not
been well analyzed.}

\subsection{The Invisible Argument}

\otter\ recognizes a built-in unary function symbol
\verb:$IGNORE(_):.
%$
Forward subsumption treats each term that starts with
\verb:$IGNORE: as the constant \verb:$IGNORE:, completely ignoring its
argument.  For example, \verb:p(a,$IGNORE(b)): subsumes
\verb:p(a,$IGNORE(c)):.
All other operations (in particular, inference rules, demodulation, and
back subsumption) treat \verb:$IGNORE: as an ordinary function symbol.

The purpose of \verb:$IGNORE: is to record data about the derivation
of a clause without having that data prevent the forward subsumption
of clauses that would be subsumed without that data.  The
\verb:$IGNORE: term is the term analog of the answer literal.  For
example, one can use \verb:$IGNORE: terms in the jugs and water puzzle
(Sec.~\ref{eval-examples}) to record the sequence of pourings that leads to
each state.

\subsection{Floating-Point Operations} \label{float}

Table \ref{float-tab} lists a set of floating-point evaluable
functions and predicates that are analogous to the integer arithmetic
operations listed in Sec.~\ref{eval}.  They operate in the
same way as the integer operations.
\begin{table}[ht] \centering
\caption{Floating-Point Operations}  \label{float-tab}
\begin{tabular}{|l|l|}
\hline
$float \times float \rightarrow float$  & \verb:$FSUM:, \verb:$FPROD:, \verb:$FDIFF:, \verb:$FDIV: \\
\hline
$float \times float \rightarrow bool$  & \verb:$FEQ:, \verb:$FNE:, \verb:$FLT:, \verb:$FLE:, \verb:$FGT:, \verb:$FGE: \\
\hline
\end{tabular}
\end{table}

The floating-point constants, however, are a little peculiar, both in
the way they look and in the way they behave.  They are written as
quoted strings, using either single or double quotes.  (Otherwise,
they would not be able to contain decimal points.)  Other than the
quotation marks, the form of the floating-point constants accepted by
\otter\ is exactly the same as the form accepted by the C programming
language (actually the C library used by the compiler).  Examples are
\verb:"1.2":, \verb:"10e6":, \verb:"-3.333E-5":.  A floating-point
constant must contain either a decimal point or an exponent character
\verb:e: or \verb:E:.

The peculiar behavior comes from the fact \otter\ stores the floating
point numbers as character strings instead of directly as floating
point numbers.  To apply a floating-point operation, \otter\ starts
with the operand strings, translates them to true floating-point
numbers (the C data type ``double'' is used), performs the operation,
then translates the result into a string so that it can be an \otter\
constant.  As well as being inefficient, this scheme also has a
problem with precision, because a fixed format is used to translate
the results back into strings.  The default format is \verb:"%.12f":,
and it can be changed with a command such as
\begin{verse} \small
\verb:float_format("%17.8f"):
\end{verse}
{\em Caution.}  \otter\ does not check that the string
in the \verb:float_format: command is a well-formed format
specification.  This is the user's responsibility.

To fully understand how this
works, see the standard C language reference \cite[Appendix B]{c-2ed};
in particular, the C library functions \verb:sscanf: and
\verb:sprintf: are used to translate to and from strings.

\subsection{Foreign Evaluable Functions} \label{foreign}

\otter\ provides a general mechanism through which one can create one's
own evaluable functions and predicates.  The user (1) declares the
function, its argument types, and its result type, (2) inserts a call to the
function in the \otter\ source code, (3) writes a C routine to implement
the function, and (4) recompiles \otter.
The user must have a personal copy of the source code to use this
feature.  See the source code file \verb:foreign.h: for step-by-step
instructions, examples, templates, and test files.

\noindent {\em Important note.}   Many times you can avoid having
to do all of this by just writing your function with demodulators and
using existing built-in functions.  For example, if you need the maximum
of two doubles, you can just use the demodulator
\verb:float_max(x,y) = $IF($FGT(x,y), x, y):.

\subsection{The Inference Rule $gL$ for Cubic Curves} \label{gl}

Based on work of R.~Padmanabhan and others, a new inference rule, $gL$
(``geometric Law'', or ``Local to global''), was added to \otter.  The rule
implements a local-to-global generalization principle that has a
geometric interpretation for cubic curves.  The article \cite{gl}
and the monograph \cite{wm-rp:monograph}
contain descriptions of the rule, some details about its
implementation in \otter, and several new results obtained with its use.

The rule $gL$ applies to single positive unit equalities, and it
is implemented in two ways: as an inference rule, with unification,
and as a rewrite rule, for when the target terms are already identical.

Flag \noindent \verb:geometric_rule:.  Default clear.  When this flag
is set, $gL$ is applied as an inference rule (along with any other
inference rules that are set) to each given clause.  The rule $gL$
applies to single positive unit equalities.

Flag \noindent \verb:geometric_rewrite_before:.  Default clear.  When this flag
is set, $gL$ is applied as a rewrite rule, before ordinary demodulation,
to each generated clause.

Flag \noindent \verb:geometric_rewrite_after:.  Default clear.  When this flag
is set, $gL$ is applied as a rewrite rule, after ordinary demodulation,
to each generated clause.

Flag \noindent \verb:gl_demod:.  Default clear.  When this flag
is set, ordinary demodulation is not applied to any derived clauses.
Instead, after a clause is kept, it is copied, and the copy
is demodulated and processed.

Our experience has shown that given two equalities of equal weight,
one the result of $gL$ and the other not, the $gL$ result is usually
more interesting.  The following parameter can give preference to
$gL$ results.

\noindent \verb:assign(geo_given_ratio,:$n$\verb:):.  Default 1, range [$-1$..\maxint ].
When this parameter is not $-1$, it affects selection of the given
clause in a way similar to
\verb:pick_given_ratio:.  If the ratio is $n$, then
for each $n$ given clauses selected in the normal way by weight,
one given clause is selected because it is the lightest $gL$ result
available in \verb:sos:.  If \verb:pick_given_ratio: and
\verb:geo_given_ratio: are both in effect, then clashes are resolved
in favor of \verb:geo_given_ratio:.

\subsection{Linked UR-Resolution}

\otter\ has an inference rule, \verb:linked_ur_res:, that is
an application of the linked inference principle to
UR-resolution.  Linked inference rules can take much
larger inference steps than the corresponding nonlinked rules,
thereby avoiding the retention of many clauses that correspond
to low-level deduction steps which can interfere with
the overall proof search strategy.

We refer the reader to \cite{link-I,link-II,link-implement}
for background on linked inference rules, and we focus
here on specifying the constraints on linked UR-resolution
for \otter.
The constraints are specified by six flags, two parameters,
and annotations on input clauses.
\newpage
\paragraph{Linked UR Flags} \strut

\noindent
Flag \verb:linked_ur_res:.  Default clear.  If this flag is set,
linked UR-resolution is applied to all given clauses.

\noindent
Flag \verb:linked_ur_trace:.  Default clear.  If this flag is set,
detailed information about the linking process is sent to the
output file.

\noindent
Flag \verb:linked_sub_unit_usable:.  Default clear.  If this flag is set,
intermediate unit clauses are checked for subsumption against
the \verb:usable: list.

\noindent
Flag \verb:linked_sub_unit_sos:.  Default clear.  If this flag is set,
intermediate unit clauses are checked for subsumption against
the \verb:sos: list.

\noindent
Flag \verb:linked_unit_del:.  Default clear.  If this flag is set,
unit deletion is applied to intermediate clauses.

\noindent
Flag \verb:linked_target_all:.  Default clear.  If this flag is set,
any literal can be a target.

\paragraph{Linked UR Parameters} \strut

\noindent
\verb:assign(max_ur_depth,:$n$\verb:):.  Default 5, range [0 .. 100].
This parameter limits the depth of linked UR-resolution.
Note that the depth of ordinary UR-resolution is 0.

\noindent
\verb:assign(max_ur_deduction_size,:$n$\verb:):.  Default 20, range [0 .. 100].
This parameter limits the size of linked UR-resolution inferences,
that is, the number of corresponding binary resolution steps.
In other words, the size of a linked inference step is one less than
the number of clauses that participate.

\paragraph{Linked UR Annotations} \strut

\noindent
Each clause that participates in a linked UR-resolution inference
is classified as
a \emph{nucleus} (the nonunit clause containing the target literal),
a \emph{link} (nonunit clauses all of whose literals are resolved),
or a \emph{satellite} (unit clauses).

Input clauses can be annotated with special literals
specifying the role(s) they can play in linked UR inferences.
The clause annotations are as follows.

\noindent
\verb:$NUCLEUS([:\emph{list-of-literal-numbers}\verb:]): ---
%$
the clause (assumed to be nonunit) can be a nucleus.
The argument is a list of positive integers identifying the
literals that can act as targets.

\noindent
\verb:$LINK([]): ---
%$
the clause (assumed to be nonunit) can act as a link.
The argument must be the empty list.

\noindent
\verb:$BOTH([:\emph{list-of-literal-numbers}\verb:]): ---
%$
the clause (assumed to be nonunit) can be either a nucleus or a link,
and when it is used as a nucleus, the admissible target literals
are given in the list.

\noindent
\verb:$SATELLITE([]): ---
%$
the clause (assumed to be unit) can act as a satellite.
The argument must be the empty list.

For example, the annotation on the following four-literal clause says that
it can act as a nucleus with the fourth literal as the target.

{\small
\begin{verbatim}
    $NUCLEUS([4]) | -go | -P31 | -Q31 | R3_LD1_DS5.
\end{verbatim}
}
%$
\noindent
Input clauses on the \verb:usable: list must be annotated
to participate in linked UR.  Units on the list \verb:sos:
are assumed to be satellites and need not be annotated.

Most experiments with linked UR-resolution have been done
under the following constraints.  (1) Linked UR is the only
inference rule being used, (2) every input clause in the
\verb:usable: list is annotated, and (3) every clause
in the \verb:sos: list is a unit and is \emph{not}
annotated.  Linked UR seems to behave correctly under
these constraints, but several problems have been noticed
with other initial conditions.

\paragraph{Acknowledgment.}
The linked UR-resolution rule was implemented by Nick Karonis,
with collaboration from Bob Veroff and Larry Wos.

\subsection{Splitting}

To address \otter's poor performance on many non-Horn
problems, a splitting rule was installed in \otter\ (in November 1997).
By ``splitting'' we mean that the search is divided into
two or more independent branches such that if each of the branches
is refuted, then the state before the split has been refuted.
Splitting is typically recursive.

\otter's splitting implementation uses the \unix\ fork() system
call, which creates copies of the state of the \otter\ process.
An additional hypothesis is asserted on the first branch,
and the first branch continues executing while the second branch
waits.  If the first branch is refuted, the second branch starts
running with its additional hypothesis.
This method avoids explicit backtracking.

Two splitting methods are available: splitting on ground clauses,
and splitting on ground atoms.  In both methods, the parameter
\verb:split_depth: can be used to limit the depth of splitting.
For example, with the command

{\small
\begin{verbatim}
    assign(split_depth, 3).
\end{verbatim}
}
\noindent
a case such as [1.1.1.1] will not occur.

\subsubsection{Splitting on Ground Clauses}

Clause splitting can be triggered in two ways: either periodically
or when a ground clause is selected as the given clause.
In both methods, the clauses on which splitting occurs can
be constrained by any of the following three flags (all clear by default).

{\small
\begin{verbatim}
    set(split_pos).     % split on positive clauses only
    set(split_neg).     % split on negative clauses only
    set(split_nonhorn). % split on non-Horn clauses only
\end{verbatim}
}
\noindent
These flags determine eligibility.
If none of the flags is set, all ground nonunit clauses are eligible.

\paragraph{Splitting Periodically on Clauses.}  To enable the periodic
splitting method, one uses the following command.

{\small
\begin{verbatim}
    set(split_clause).
\end{verbatim}
}
\noindent
The default period is every 5 given clauses.  To change the period,
say to 10 given clauses, use the following command.

{\small
\begin{verbatim}
    assign(split_given, 10).
\end{verbatim}
}
\noindent
Instead of splitting after some number of given clauses, one
can split after some number of seconds, say 4, with the following command.

{\small
\begin{verbatim}
    assign(split_seconds, 4).
\end{verbatim}
}
\noindent
The clause on which to split can be selected from the set of eligible
clauses in two ways.  The default method is to select the first, lightest
(using the pick-given scale) eligible clause from the sequence
\verb:usable+sos:.  Instead, one can use the command

{\small
\begin{verbatim}
    set(split_min_max).
\end{verbatim}
}
\noindent
which says to use the following method to compare two eligible clauses.
Prefer the clause with the lighter heaviest literal
(using the pick-given scale);  if the heaviest literals have the same
weight, use the lighter clause;  if the clauses have the same
weight, use the first in \verb:usable+sos:.

\paragraph{Splitting When Given.}  To specify that clause splitting should
be occur whenever an eligible clause is selected as the given clause, one uses
the following command.

{\small
\begin{verbatim}
    set(split_when_given).
\end{verbatim}
}

\paragraph{The Branches for Clause Splitting.}
If \otter\ decides to split on a clause, say \verb:P|Q|R:,
the assumptions for the three cases are

{\small
\begin{verbatim}
    Case 1: P.
    Case 2: -P & Q.
    Case 3: -P & -Q & R.
\end{verbatim}
}
\noindent
One system fork occurs, and Case 1 executes.  If it succeeds,
a second fork occurs, and Case 2 executes.  If that succeeds,
Case 3 executes.  If any of the cases fails to produce a
refutation, the failure is propagated to the top, and the entire
search fails.

\subsubsection{Splitting on Atoms}

To split on atoms, \otter\ periodically selects a ground
atom, say \verb:P:, and considers two branches, one with
assertion \verb:P:, and the other with \verb:-P:.  The following
command specifies splitting on atoms.

{\small
\begin{verbatim}
    set(split_atom).
\end{verbatim}
}
\noindent
The parameters \verb:split_given: and \verb:split_seconds:
determine (just as for clause splitting) when atom splitting
occurs.
If all input clauses are ground, and if the parameter
\verb:split_given: is assigned 0, then the resulting procedure
is essentially a (very slow) Davis-Putnam-Loveland-Logemann SAT procedure.

An atom is eligible for splitting if it occurs in
an eligible nonunit ground clause.  Clause eligibility
is determined just as in the clause splitting case, that
is, by the flags \verb:split_pos:, \verb:split_neg:, and
\verb:split_nonhorn:.

All clauses in \verb:usable+sos: are considered when deciding
the best eligible atom.
The default method select the
lightest atom in the lightest clause (using the pick-given scale).  An
optional method for selecting an atom considers the number of
occurrences of the atom.  The command

{\small
\begin{verbatim}
    set(split_popular).
\end{verbatim}
}
\noindent
says to prefer the atom that occurs in the greatest number
of clauses.

Instead of having \otter\ decide the atoms on which to split,
the user can specify them in the input file with a command such as

{\small
\begin{verbatim}
    split_atoms([P, a=b, R]).
\end{verbatim}
}
\noindent
which says to split, in order, on those atoms.  In this example,
we get eight cases, and then no more splitting occurs.  The
time at which the splitting occurs is determined, as above,
by the parameters \verb:split_given: and \verb:split_seconds:.

\subsubsection{More on Splitting}

If \otter\ fails to find a proof for a particular case (e.g., the list
\verb:sos: empties or some limit is reached), the whole attempt fails.  If
the search strategy is complete, then an empty \verb:sos: list indicates
satisfiability, and the set of assumptions introduced by splitting
give a partial model.  It is up to the user, however, to complete
the model.

When \otter\ finds a refutation by splitting, the output file
does not contain an overall proof.  A proof is given for
each leaf in the tree, and those proofs contain clauses such as

{\small
\begin{verbatim}
    496 [264,split.1.1.1.1] nop(C,D)!=nop(A,A).
\end{verbatim}
}
\noindent
in which the justification indicates that a split occurred on
clause 264, and this clause is the assertion for case [1.1.1.1].
Other information about splitting is given in the output file,
for example, when a split occurs, the case numbers, the
case assertions, and when forked processed begin and end.

When splitting is enabled, the parameter \verb:max_seconds:
(for the initial process and all descendant processes) is checked
against the wall clock (from the start of the initial process)
instead of against the process clock.
This is problematic if the computer is busy with other processes.

\otter's splitting rule is highly experimental, and we do not
have much experience with it.
A general strategy that may be useful for non-Horn problems is the following.

{\small
\begin{verbatim}
    set(split_when_given).
    set(split_pos).            % Also try it without this command.
    assign(split_depth, 10).
\end{verbatim}
}
\noindent
The \otter\ distribution packages should contain a directory
of sample input files that use the splitting rule.

\paragraph{Acknowledgment}
The splitting rule was developed in collaboration with
Dale Myers, Rusty Lusk, and Mohammed Almulla.

% end of fringe features

\section{Soundness and Completeness}

\subsection{Soundness} \label{ivy}

\otter\ has a very good record with respect to soundness,
but (as far as we know) no parts of it (the C code) have been formally
verified.  If anything depends on proofs found by
\otter, the proofs should be carefully checked, by hand or
by an independent program.

The IVY project \cite{ivy} contains a component that
checks the proof objects (Sec.~\ref{output-flags})
produced by \otter.  The main result of the IVY project is
a hybrid system, constructed in the ACL2 verification
environment \cite{acl2-approach}, that takes a first-order
conjecture, translates it to clauses, sends the clauses
to \otter, and checks any proof objects that are returned.
ACL2 has been used to prove various soundness properties of
the clause translator, the proof checker, and their
composition as a hybrid system.

\subsection{Completeness}

If the clause set does not involve equality, or if it involves
equality and includes the equality axioms, then many of the common
refutation-complete resolution search strategies can be easily
achieved with \otter.  For example, hyperresolution and factoring,
with positive clauses in the list \verb:sos: and nonpositive clauses in
the list \verb:usable:, is complete.  If the input clause set is Horn,
then factoring is not required.
The default method of selecting the given clause (take one with the
fewest symbols) does not interfere with completeness, and neither
forward nor back subsumption, as implemented in
\otter, interferes with completeness of the basic inference
rules.

Completeness issues are more complex when paramodulation is the
inference rule, especially when the set of support strategy is
considered.  A simple and complete paramodulation strategy for \otter\
is (1) paramodulate {\em from} and {\em into} the given clause, (2)
paramodulate {\em from} and {\em into} both sides of equality
literals, (3) paramodulate {\em from} (but not {\em into}) variables,
and (4) place all input clauses in the list \verb:sos:.  The equality
\verb:x=x: is required, but the functionally reflexive axioms are not
required.

Completeness of the basic inference rules is important, but incomplete
restrictions and refinements are frequently required to find proofs.
For example, we almost always use the \verb:max_weight: parameter;
strictly speaking, it is incomplete, but it saves a lot of time and
memory, and careful use of it does not prevent \otter\ from finding
proofs in practice.  For paramodulation, we generally use the
flag \verb:anl_eq: with additional restrictions---some
are known to be incomplete, and others have not been analyzed.
We sometimes use UR-resolution on non-Horn sets, which is incomplete.
And we make extensive use of weighting to purge uninteresting
clauses and the options \verb:delete_identical_nested_skolem:,
\verb:max_distinct_vars:, and \verb:max_literals:, all of which
interfere with completeness.

\section{Limits, Abnormal Ends, and Fixes}

\otter\ has several compile-time limits.
If a limit is exceeded, a message containing the name of the limit
will appear in the output file and/or at the terminal.  To raise the
limit, find the appropriate definition (\verb:#define:) in a
\verb:.h: or \verb:.c: file, increase the limit, and recompile \otter.
(Of course, one must have one's own copy of the source code to
do this.)  Some of the limits are as follows.

\noindent \verb:MAX_NAME: --- Maximum number of characters in a variable,
constant, function, or predicate symbol.

\noindent \verb:MAX_BUF: --- Maximum number of characters in an input string
(clause, formula, command, weight template, etc.).

\noindent \verb:MAX_VARS: --- Maximum number of distinct variables in a clause.

\noindent \verb:MAX_FS_TERM_DEPTH: --- Maximum depth of terms in
the forward subsumption discrimination tree.

\noindent \verb:MAX_AL_TERM_DEPTH: --- Maximum depth of left-hand arguments of
equalities in the demodulation discrimination tree.

\paragraph{Conserving Memory.}
Several steps can be taken if \otter\ is using too much memory.
\begin{itemize}
\item
Use \verb:max_weight: to discard (more) generated clauses.
This is a very effective way to save memory (and time).
\item
Set the flag \verb:control_memory: (Sec.~\ref{misc-flags}), or use the parameters \\
\verb:change_limit_after: and \verb:new_max_weight: (Sec.~\ref{mem-control}).
\item
Decrease (down to 0) the value of the \verb:fpa_literals: and
\verb:fpa_terms: parameters.
\item
Set the \verb:for_sub_fpa: flag to switch forward subsumption indexing from
discrimination tree to {\sc fpa} indexing.
\item
If the inference rules being used are binary resolution or
paramodulation, clear the flag \verb:detailed_history:.
\item
If a lot of back subsumption or back demodulation is expected,
set the flag \verb:really_delete_clauses: (Sec.~\ref{misc-flags}).
\item
If applicable, set \verb:no_fapl: or \verb:no_fanl:
(Sec.~\ref{index-flags}).
\item
If back demodulation is being used, clear the flag \verb:index_for_back_demod:.
\item
Run an \otter\ job until memory runs out, collect interesting
lemmas from the output file, then rerun the job including the
lemmas as input clauses.  Repeat.  (This can be a good strategy
even when memory is not a problem.)
\end{itemize}

\section{Obtaining and Installing \otter}

\otter\ 3 is free, and there are no restrictions on copying
or distributing it.  The main means of distribution is from
the \otter\ Web site at \verb|http://www.mcs.anl.gov/AR/otter/|.

Once one has a copy of the \otter\ 3 distribution directory,
one should look at the file \verb:README: for instructions on installing
and testing \otter.  On \unix-like systems, \otter\ may
have to be compiled.  There may also be executable versions for
Microsoft Windows available on the \otter\ Web site.

\addcontentsline{toc}{section}{Acknowledgments}
\section*{Acknowledgments}

Much of my work over the
past few years has been in collaboration with Larry Wos.  Toward our
goal of creating programs that are expert assistants for
mathematicians, logicians, engineers, and other scientists, we have worked
together on many applications of automated deduction, and that work
has led to many of \otter's current features.

The basic design of the program, including the data structures and the
use of indexing, descends mostly from theorem provers designed and
implemented by Ross Overbeek.
The indexing mechanisms, which are in large part responsible for the
performance of the program, have benefited from discussions with
Overbeek, Mark Stickel, and Rusty Lusk.

For many years Bob Veroff has maintained his own versions of \otter.
Many of his enhancements have been adopted for the official versions
of \otter.

The expert users of \otter, including
Larry Wos,
Bob Veroff,
John Kalman,
Ken Kunen,
Art Quaife,
Dale Myers,
Johan Belinfante,
Michael Beeson,
Branden Fitelson, and
Zac Ernst,
have tracked down bugs and suggested useful enhancements.
\newpage
\addcontentsline{toc}{section}{References}
\bibliographystyle{plain}

% \bibliography{/home/mccune/papers/bib/master}

\begin{thebibliography}{10}

\bibitem{after-25-years}
W.~Bledsoe and D.~Loveland, editors.
\newblock {\em Automated Theorem Proving: After 25 Years}, volume~29 of {\em
  Contemporary Mathematics}.
\newblock AMS, 1984.

\bibitem{boyer-moore-2}
R.~S. Boyer and J~S. Moore.
\newblock {\em A Computational Logic Handbook}.
\newblock Academic Press, New York, 1988.

\bibitem{chang-lee}
C.-L. Chang and R.~C.-T. Lee.
\newblock {\em Symbolic Logic and Mechanical Theorem Proving}.
\newblock Academic Press, New York, 1973.

\bibitem{termination}
N.~Dershowitz.
\newblock Termination of rewriting.
\newblock {\em J. Symbolic Computation}, 3:69--116, 1987.

\bibitem{hw-verify}
C.~A.~R. Hoare and M.~J.~C. Gordon, editors.
\newblock {\em Mechanized Reasoning and Hardware Design}.
\newblock Prentice Hall, 1992.

\bibitem{bm-verify}
{J S. Moore, editor}.
\newblock Special {I}ssue on {S}ystem {V}erification.
\newblock {\em J. Automated Reasoning}, 5(4), 1989.

\bibitem{rta-85}
J.-P. Jouannaud, editor.
\newblock {\em Rewriting Techniques and Applications, Lecture Notes in Computer
  Science, Vol. 202}, Berlin, 1985. Springer-Verlag.

\bibitem{kalman-otter}
J.~Kalman.
\newblock {\em Automated Reasoning with Otter}.
\newblock Rinton Press, Princeton, New Jersey, 2001.

\bibitem{rrl}
D.~Kapur and H.~Zhang.
\newblock {RRL}: {R}ewrite {R}ule {L}aboratory {U}ser's {M}anual.
\newblock Technical Report 89-03, Department of Computer Science, University of
  Iowa, 1989.

\bibitem{acl2-approach}
M.~Kaufmann, P.~Manolios, and J~Moore.
\newblock {\em Computer-Aided Reasoning: An Approach}.
\newblock Advances in Formal Methods. Kluwer Academic, 2000.

\bibitem{c-2ed}
B.~Kernighan and D.~Ritchie.
\newblock {\em The C Programming Language}.
\newblock Prentice Hall, second edition edition, 1988.

\bibitem{knuth-bendix}
D.~Knuth and P.~Bendix.
\newblock Simple word problems in universal algebras.
\newblock In J.~Leech, editor, {\em Computational Problems in Abstract
  Algebras}, pages 263--297. Pergamon Press, Oxford, 1970.

\bibitem{kurosh}
A.~G. Kurosh.
\newblock {\em The Theory of Groups}, volume~1.
\newblock Chelsea, New York, 1956.

\bibitem{loveland}
D.~Loveland.
\newblock {\em Automated Theorem Proving: A Logical Basis}.
\newblock North-Holland, Amsterdam, 1978.

\bibitem{ITP}
E.~Lusk and R.~Overbeek.
\newblock The {A}utomated {R}easoning {S}ystem {ITP}.
\newblock Tech. Report ANL-84/27, Argonne National Laboratory, Argonne, IL,
  April 1984.

\bibitem{skolem-aaai}
W.~McCune.
\newblock Skolem functions and equality in automated deduction.
\newblock In {\em Proceedings of the Eighth National Conference on Artificial
  Intelligence}, pages 246--251, Cambridge, MA, 1990. MIT Press.

\bibitem{mace2}
W.~McCune.
\newblock {MACE} 2.0 {R}eference {M}anual and {G}uide.
\newblock Tech. Memo ANL/MCS-TM-249, Mathematics and Computer Science Division,
  Argonne National Laboratory, Argonne, IL, June 2001.

\bibitem{wm-rp:monograph}
W.~McCune and R.~Padmanabhan.
\newblock {\em Automated Deduction in Equational Logic and Cubic Curves},
  volume 1095 of {\em Lecture Notes in Computer Science (AI subseries)}.
\newblock Springer-Verlag, Berlin, 1996.

\bibitem{ivy}
W.~McCune and O.~Shumsky.
\newblock {IVY}: A preprocessor and proof checker for first-order logic.
\newblock In M.~Kaufmann, P.~Manolios, and J~Moore, editors, {\em
  Computer-Aided Reasoning: ACL2 Case Studies}, chapter~16. Kluwer Academic,
  2000.

\bibitem{gl}
R.~Padmanabhan and W.~McCune.
\newblock Automated reasoning about cubic curves.
\newblock {\em Computers and Mathematics with Applications}, 29(2):17--26,
  1995.

\bibitem{quaife-geo}
A.~Quaife.
\newblock Automated development of {T}arski's geometry.
\newblock {\em J. Automated Reasoning}, 5(1):97--118, 1989.

\bibitem{quaife-thesis}
A.~Quaife.
\newblock {\em Automated Development of Fundamental Mathematical Theories}.
\newblock PhD thesis, University of California at Berkeley, 1990.

\bibitem{siekmann-wrightson}
J.~Siekmann and G.~Wrightson, editors.
\newblock {\em Automation of Reasoning: Classical Papers on Computational
  Logic}, volume 1 and 2.
\newblock Springer-Verlag, Berlin, 1983.

\bibitem{AURA}
B.~Smith.
\newblock Reference {M}anual for the {E}nvironmental {T}heorem {P}rover: {A}n
  {I}ncarnation of {AURA}.
\newblock Tech. Report ANL-88-2, Argonne National Laboratory, Argonne, IL,
  March 1988.

\bibitem{tptp-web}
G.~Sutcliffe and C~Suttner.
\newblock The {TPTP} {P}roblem {L}ibrary for {A}utomated {T}heorem {P}roving.
\newblock \verb|http://www.tptp.org/|.

\bibitem{link-implement}
R.~Veroff.
\newblock An {A}lgorithm for the {E}fficient {I}mplementation of {L}inked
  {UR}-resolution.
\newblock Tech. Report CD92-17, Department of Computer Science, University of
  New Mexico, 1992.

\bibitem{veroff:hints}
R.~Veroff.
\newblock Using hints to increase the effectiveness of an automated reasoning
  program: Case studies.
\newblock {\em J. Automated Reasoning}, 16(3):223--239, 1996.

\bibitem{veroff:sketches}
R.~Veroff.
\newblock Solving open questions and other challenge problems using proof
  sketches.
\newblock {\em J. Automated Reasoning}, 27(2):157--174, 2001.

\bibitem{link-I}
R.~Veroff and L.~Wos.
\newblock The linked inference principle, i: The formal treatment.
\newblock {\em J. Automated Reasoning}, 8(2):213--274, 1992.

\bibitem{book2}
L.~Wos.
\newblock {\em Automated Reasoning: 33 Basic Research Problems}.
\newblock Prentice-Hall, Englewood Cliffs, NJ, 1988.

\bibitem{book1a}
L.~Wos, R.~Overbeek, E.~Lusk, and J.~Boyle.
\newblock {\em Automated Reasoning: Introduction and Applications, {\rm 2nd
  edition}}.
\newblock McGraw-Hill, New York, 1992.

\bibitem{JAR-overview}
L.~Wos, F.~Pereira, R.~Boyer, J~Moore, W.~Bledsoe, L.~Henschen, B.~Buchanan,
  G.~Wrightson, and C.~Green.
\newblock An overview of automated reasoning and related fields.
\newblock {\em J. Automated Reasoning}, 1(1):5--48, 1985.

\bibitem{fascinating}
L.~Wos and G.~Pieper.
\newblock {\em A Fascinating Country in the World of Computing: Your Guide to
  Automated Reasoning}.
\newblock World Scientific, Singapore, 1999.

\bibitem{link-II}
L.~Wos, R.~Veroff, B.~Smith, and W.~McCune.
\newblock The linked inference principle {II}: {T}he user's view.
\newblock In R.~Shostak, editor, {\em Proceedings of the 7th International
  Conference on Automated Deduction, Lecture Notes in Computer Science, Vol.
  170}, pages 316--332, Berlin, 1984. Springer-Verlag.

\end{thebibliography}

\end{document}